\documentclass[12pt,a4paper]{article}
\usepackage[polish,english]{babel}
\usepackage[utf8]{inputenc}
\usepackage[T1]{fontenc}
\usepackage{authblk} % authors and affiliations
\let\babellll\lll
\let\lll\relax
\usepackage{amsmath}
\usepackage{amsfonts}
\usepackage{amssymb} 
\let\lll\babellll % \lll z babel przywrocone jako \lll
\usepackage{graphicx}
\usepackage{epstopdf}
\usepackage{color}
\newcommand{\RNumb}{\mathbb{R}}

\newcommand{\ZNumb}{\mathbb{Z}}

\newcommand{\UnitOp}{1\kern-4.5pt 1} % matrix unit
\newcommand{\MatUnit}{1\kern-3pt 1} % matrix unit
\newcommand{\Trace}{\mathrm{Tr}} % trace
\newcommand{\Prob}{\mathrm{Prob}\,} % probability
\newcommand{\Group}[1]{\textrm{#1}} % group assignment
\newcommand{\Bra}[1]{\langle #1 \vert} % bra
\newcommand{\Ket}[1]{\vert #1 \rangle} % ket
\newcommand{\BraKet}[2]{\langle #1 \vert #2 \rangle} % bra(c)ket
\newcommand{\Aver}[1]{\langle #1 \rangle}
\newcommand{\EqCond}[1]{\overset{#1}{=}}
\newcommand{\Var}{\mathrm{var}}
\newcommand{\Ham}{\hat{\mathcal{H}}} % Hamiltonian

\newcommand{\EOp}{\mathsf{E}\kern-1pt\llap{$\vert$}}   % PEv E operator
\newcommand{\WOp}{\hat{\mathsf{W}\kern-1pt\llap{$-$}}} % PEv W generator 
\newcommand{\FOp}{\mathsf{F}\kern-1pt\llap{$\vert$}}   % QEv F operator  
\newcommand{\pev}[1]{\mathrm{pev}\kern-2pt\left( #1 \right)}
\newcommand{\ro}[1]{\rho(\tau_{#1};\nu_{#1})}
\newcommand{\roprime}[1]{\rho'(\tau_{#1};\nu_{#1})}
\newcommand{\ronk}{\rho(\tau_n;\nu_{n})}
\newcommand{\ronl}{\rho(\tau_{n-1};\nu_{n-1})}
\newcommand{\StateSpace}[1]{\mathcal{#1}} % state space
\newcommand{\SpaceTime}[1]{\mathbf{#1}} % space--time

\newcommand{\Komentarz}[1]{} % comment

\begin{document}
%%%%%%%%%%%%%%%%%%%%%%%%%%%%%%%%%%%%%%%%%%%%%%%%%%%%%%%%%%%%%%%%%%%%%%%%

\title{Projection evolution and quantum spacetime}

%%%%%%%%%%%%%%%%%%%%%%%%%%%%%%%%%%%%%%%%%%%%%%%%%%%%%%%%%%%%%%%%%%%%%

\author{A.~G\'o\'zd\'z 
\thanks{email: Andrzej.Gozdz@umcs.lublin.pl, ORCID: 0000-0003-4489-5136}}
\affil{Institute of Physics,
Maria Curie--Sk\l{}odowska University in Lublin, Poland}

\author{M.~G\'o\'zd\'z 
\thanks{email: mgozdz@kft.umcs.lublin.pl, ORCID: 0000-0003-4958-8880}}
\affil{Institute of Computer Science,
Maria Curie--Sk\l{}odowska University in Lublin, Poland}

\author{A.~P{\c e}drak 
\thanks{email: Aleksandra.Pedrak@ncbj.gov.pl, ORCID: 0000-0002-8808-3239}}
\affil{National Centre for Nuclear Research, Warsaw, Poland}

\date{\today}

%%%%%%%%%%%%%%%%%%%%%%%%%%%%%%%%%%%%%%%%%%%%%%%%%%%%%%%%%%%%%%%%%%
\maketitle
%%%%%%%%%%%%%%%%%%%%%%%%%%%%%%%%%%%%%%%%%%%%%%%%%%%%%%%%%%%%%%%%%%

\begin{abstract}
We discuss the problem of time in quantum mechanics. In the traditional
formulation time enters the model as a~parameter, not an observable. In our
model time is a~quantum observable as any other quantum quantity and it is also
a component of the spacetime position operator. In this case, instead of the
unitary time evolution, other operators, usually projection or POVM operators
which map the space of initial states into the space of final states at each
step of the evolution can be used. The quantum evolution itself is a stochastic
process. This allows to treat time as a~quantum observable in a~consistent,
observer independent way, which is a very important feature to resolve some
quantum paradoxes and the time problem in cosmology.

An idea of construction of a quantum spacetime as a special set of the allowed
states is presented. An example of a structureless quantum Minkowski--like
spacetime is also considered.

We present the projection evolution model and show how the traditional
Schr\"odinger evolution and relativistic equations can be obtained from it, in
the flat structureless spacetime.

We propose the form of the time operator which satisfies the energy-time
uncertainty relation based on the same inequality as the space position and
spatial momenta observables.  The sign of the temporal component of the
four-momentum operator defines the basic arrow of time in spacetime.
    
\end{abstract}

% 03.65.-w   quantum mechanics
% 03.65.Ca   foundations of quantum mechanics, measurement theory
% 03.65.Ta   formalism
% \pacs{03.65.-w, 03.65.Ca, 03.65.Ta}

% \noindent {\it Keywords\/}: quantum mechanics, time operator, time
% evolution, interference in time
%%%%%%%%%%%%%%%%%%%%%%%%%%%%%%%%%%%%%%%%%%%%%%%%%%%%%%%%%%%%%%%%%%%%%%
\section{Introduction}
%%%%%%%%%%%%%%%%%%%%%%%%%%%%%%%%%%%%%%%%%%%%%%%%%%%%%%%%%%%%%%%%%%%%%%

For many years time was treated in physics as a~universal parameter
which allows the observer to divide the reality into past, present, and
future. What is more, time was flowing always in one direction, called the
arrow of time. This direction implied also the direction of changes that
may spontaneously happen to any physical system, which ultimately leads
to the notion of causality. We are used to the fact that past affects
future, but future cannot affect the past, as this will act against the
arrow of time.

The development of relativity theory changed this picture in a~substantial
way. To obtain a realistic model one needs to treat time and space in a~way
consistent with the relativity theory.  The metric and other tensors gained
their time components which were transforming during the change of the
coordinate system along with the spatial coordinates. For example, the position
and the linear momentum are four-vectors $x^\mu$ and $p^\mu$, $\mu=0,1,2,3$.
They take the form $x^\mu=(x^0,\vec x)$ and $p^\mu=(p^0,\vec p)$, where $x^0$
represents time and $p^0 = E/c$, $E$ being the total energy. This feature is
absent in the non-relativistic physics.

One may ask if time and space positions behave in the same way in the
macroscopic and microscopic scales? We know that both non-relativistic and
relativistic physics agree upon the basic properties of time, so if one expects
any deviations from the standard picture, one should look at quantum
mechanics.

In the standard formulation of the quantum theory, any physical quantity is
represented by a~self adjoint operator whose eigenvalues are the possible
outcomes of its measurement. However, the so-called Pauli theorem
\cite{pauli-a,pauli-b} states, that it is impossible to construct a~self adjoint
time operator which would be canonically conjugate to the Hamiltonian. It
follows that time is not a~physical observable but is introduced as a~universal
numerical parameter. This approach is inconsistent with what we know from the
relativity theory, not to mention that it gives very limited means to discuss
quantum events in the time domain.  A~careful mathematical analysis of this
problem was presented by E.A. Galapon in Ref.~\cite{galapon}. He showed that
this problem can be overcome using weaker assumptions about the observables.

It has been extensively discussed how to introduce time as an observable in the
theory, as this affects the construction of the arrow of time and clocks (see
\cite{th01,th02,th03,th04,th05,th06,th07,th08,th09,th10,th11,th12,th13,th14,th15,th16,th17,th18,th19,th20,th21,th22,th-pre01,th-pre02,th-pre03,th-pre04,th-pre05,th-pre06,th-rel-mech-a,th-rel-mech-b,th-montevideo}
for recent developments). Related topics include also the problem of time in
entangled systems, the time of decoherence and the role of the energy-time
uncertainty relation. The process of quantization can be performed in different
ways and tested using specially designed experiments, like it has been shown in
the example of the time of arrival operator \cite{toa-th,toa-exp}. Since time is
connected with the energy operator, thermodynamics of quantum processes started
to be of interest \cite{termo-th01,termo-th02}. It has already been shown that
due to the quantum correlations, heat may spontaneously flow from the colder to
the hotter subsystem \cite{termo-exp}, which is not observed in the macroscopic
scale. Entanglement and the immediate change of state of both entangled
particles rises also the question how to describe \cite{ent-th01,ent-th02} and
experimentally investigate \cite{ent-exp} this process.

The problem of time appears also in systems performing quantum
computation. Most quantum protocols assume that we can neglect the time
delays introduced by quantum gates and connections in the system, which
does not have to be the case. Another problem arises with the
theoretically proposed quantum gates with feedback
\cite{qcomp01,qcomp02,qcomp03,qcomp04}, which are impossible to describe
using standard tools. Understanding the time structure of quantum
operations is also vital for constructing future quantum neural networks
\cite{qneural}.

Treating time as an observable leads to the problem of time measurements
\cite{tmeas}, also in the context of quantum cosmology
\cite{tmeas-cosmo}. As time becomes a~variable, new phenomena start to
be possible, like dark matter described by fields evolving backwards in
time \cite{dmat}.

The important role of time in quantum theories is suggested by some
experiments. In Refs.~\cite{wheeler-a,wheeler-b} J.A.~Wheeler proposed
a~Gedankenexperiment, the so called ``delayed choice problem''. This
idea has been experimentally tested by the group of A.~Aspect with the
primary intention to test Bell's inequalities \cite{exp-aspect} showing
that Wheeler's predictions were correct. Other groups
\cite{exp-wh-a,exp-wh-b,exp-wh-c,exp-wh-d,exp-wh-e,manning} arrived at
similar results. In order to investigate the problem further, the
quantum eraser was used \cite{exp-eraser,exp-space}.  The effect was
visible even when the changes introduced to the experimental setup led
to acausal events.

Another experiment was conducted using entangled pairs of photons
\cite{exp-lapalma-a,exp-lapalma-b} separated by 144~km.  Even though the
particles were causally disconnected, the changes made in the first
laboratory were affecting the second particle.

If time in the quantum regime should be treated as a~coordinate, and in fact
a~quantum observable, all physical objects' states have to have some ``width''
in the time direction, which is related to the energy-time (more precisely --
the temporal component of the four momentum operator versus time) uncertainty
relation. This means that it should be possible to observe the interference of
quantum objects through their overlap in time
\cite{exp-2slits-a,th-2slits-a,th-2slits-b,exp-2slits-b}.  It also means, that
time cannot be treated as a~parameter.
 
It seems to be very difficult to answer the fundamental question: {\it What is
  time?} An interesting hypothesis is presented in Ref.~\cite{time-ent} in which
the authors propose, that time is a~consequence of the entanglement between
particles in the universe.

Other proposal for introducing the quantuntum time is the relational quantum
mechanics summarized in Ref.~\cite{th-rel-mech-a,th-rel-mech-b,th-montevideo},
see also references therein.

In this paper we present a~consistent formulation of the quantum theory in which
the spacetime can emerge from a set of observables supported by the quantum
state space. It can be generated from a set of self-conjugated operators or
operator valued measures having expected properties. As a byproduct, such
construction should allow to obtain the time operator and the canonically
conjugated observable which represents the temporal momentum.

In the PEv approach the evolution of quantum states has to be reformulated, as
time, being a~coordinate, cannot act as a~universal ordering parameter any
longer. However, we show that the traditional time evolution, like the
Schr\"odinger, Klein-Gordon, Dirac and other equations of motion can be obtained
as special cases within our model. The problem of symmetries and conservation
laws during the evolution is also shortly discussed.

The Projection Evolution approach (PEv) is covariant, it does not need
any external observer and similarly to relational dynamics no background
is required. This implies that PEv formalism can also be applied in a
natural way to quantum gravity.

%The projection evolution approach has already its history. A~few examples of
%processes analyzed in terms of time as a quantum observable can be found in
%\cite{PEvFission2006,ArrivalTime2006,ProtonEmission2007,TimeInterf2008,DealyedC%hoice2008,ArrivalTime2008,DecayExtClock2017,NeutrinoMix2017,DelayedChoice2018}.
%The cited papers show some steps in the historical development of the PEv idea.

%%%%%%%%%%%%%%%%%%%%%%%%%%%%%%%%%%%%%%%%%%%%%%%%%%%%%%%%%%%%%%%%%%%%%%%%
\section{Projection evolution of quantum systems}
%%%%%%%%%%%%%%%%%%%%%%%%%%%%%%%%%%%%%%%%%%%%%%%%%%%%%%%%%%%%%%%%%%%%%%%%

In quantum mechanics each physical system is described by a set of all possible
observables which can be associated with it, see, e.g., the algebraic approach
to quantum mechanics \cite{emch1972,Landsman,Beny2020}.  The observables
themselves are represented either by self-adjoint operators or, more generally,
by the appropriate operator valued measures (sharp or POVM). In traditional
approaches to quantum mechanics time is not represented in the set of these
observables, it is considered to be a parameter. This inconsistency leads to
various quantum paradoxes and also to the time problem in gravity and cosmology;
the extensive set of references for the latter problem can be found in
Ref.~\cite{th-rel-mech-a,th-rel-mech-b,th-montevideo}.

One expects, the full set of observables of any physical system under
consideration contains a subset of the spacetime position operators and their
canonical conjugated momenta.  Among them the time observable and its canonical
conjugate is also expected.  It follows that time cannot be considered as
a~parameter which enumerates subsequent events but it has to be represented by
an operator or the appropriate operator measure similar to the position
observables. It means that different time characteristics of a given quantum
system can be calculated. In general, they are dependent on the state of this
system.

The assumption that quantum time, and generally the spacetime, is ``created'' by
changes of the Universe requires a modification of some parts of the paradigm of
science related to the causality and the ordering of quantum events.

%%%%%%%%%%%%%%%%%%%%%%%%%%%%%%%%%%%%%%%%%%%%%%%%%%%%%%%%%%%%%%%%%%%%%%%%

\subsection{The changes principle}

We start by formulating the fundamental principle of the projection
evolution approach: \\

{\it The evolution of a~system is a~random process caused by
  the spontaneous changes in the Universe.} \\

We call it \textbf{the changes principle}. It means that we treat
\textbf{the change as the primary process}, which allows to determine
time, space and spacetime in terms of quantum observables.

This is in contradiction with the usual thinking in which the existence of time
allows the changes to happen. In our approach the changes happen spontaneously,
according to the probability distribution, which is dictated by many factors
describing the Universe and in the case of the subsystems of this Universe also
their environments. It does not mean that the changes of a~quantum state are
totally stochastic, without any constraints. They are obviously not
deterministic, but because of interactions, symmetries which have to be
conserved, EPR correlations etc., they are related to each other and bound by
the rules of their behavior known from our experience.

As a consequence one may expect the existence of a~kind of
pseudo-causality based on the ordering of the quantum events, which
leads to the causality principle in the case of macroscopic physical
systems. In order to describe this property we introduce a~parameter
$\tau$ which orders quantum events. This parameter should be common for
the whole Universe. It should take values from an ordered set but it
does not need to have any metric structure. The parameter $\tau$ is not
an additional dimension of our space and it is not a~replacement of
time. It serves only to enumerate the subsequent steps of the evolution
of the Universe and any of its physical subsystems. The most natural
linearly ordered set is any subset of the real numbers.

In what follows we assume that the domain of the evolution parameter
$\tau$ is isomorphic to integers $\ZNumb$ or their subset. In this case
we can always use the notion of ``the next step of the evolution,''
which may be problematic for the real numbers. In the situation of
a~continuous or dense subset of the real numbers as the domain for
$\tau$, there are some conceptual difficulties which should be, if
needed, solved in the future.

An additional, very important feature of this approach is that this idea
does not need the spacetime as the background, it is background
independent.  Most of the physical theories constructed till now use the
spacetime as the primary object, with the dynamics built on top of
it. In other words, the projection evolution approach is a~background
free theory. In addition, it does not need any external observer as time
is an internal observable. This point is extremely important in the
quantum gravity and cosmology.

%%%%%%%%%%%%%%%%%%%%%%%%%%%%%%%%%%%%%%%%%%%%%%%%%%%%%%%%%%%%%%%%%%%%%%%%%%%%

\subsection{Projection evolution operators}

In the standard formulation of quantum physics, there are two kinds of
time evolution: (i) the unitary evolution, which is a~deterministic
evolution of the actual quantum state, and (ii) the stochastic
evolution, which takes place during a~measurement. The latter process
involves the projection of the quantum state onto the measured state and
can be described by one of the projection postulates.

There is a~common belief that every measurement process can be described
by means of the unitary evolution of a larger system. This approach
leads, however, to the known quantum measurement problems
\cite{Busch1996}.

The changes principle is incompatible with the unitary evolution, where
time is considered to be a~parameter. The idea of the changes principle
suggests the opposite scenario -- the primary evolution is the
stochastic evolution offered by a projection postulate. In the
projection evolution formalism we propose to use the generalized form of
the L\"uders \cite{Lueders} type of the projection postulate.

One needs to notice that this mechanism defines events as subsequent
steps of the evolution.

In the following, we introduce the evolution operators which are
formally responsible for the quantum evolution of a physical object. In
general these operators are different for different systems, similarly
to the Hamiltonian, which is a characteristic object for a given quantum
system. On the other hand one should, in principle, be able to
construct the projection evolution operators for the whole Universe
which will contain the operators for any smaller subsystem. It is due to
the fact that the proposed formalism does not require any external
observer and external variables for the evolution.

The projection evolution operator from the evolution step $\tau_{n-1}$ to the
evolution step $\tau_{n}$, where $n \in \ZNumb$, is a family of mappings from
the space of quantum states at the evolution step $\tau_{n-1}$ to the space of
quantum states at the evolution step $\tau_{n}$.

The appropriate state space, at the evolution step $\tau_{k}$, denoted by
$\mathcal{T}^{+}_{1}(\mathcal{K}(\tau_k))$, is assumed to be the space of trace
one, positive and self-adjoint operators acting in the Hilbert space
$\mathcal{K}(\tau_{k})$ , i.e., it is the space of quantum density operators.
Every corresponding Hilbert space $\mathcal{K}(\tau_{k})$, at the evolution step
$\tau_{k}$, is a subspace of a single global Hilbert space $\mathcal{K}_U$.

Note, that in this case, the simplest Hilbert space of a single, spinless
particle is not the space $L^2(\RNumb^3,d^3x)$ but the space
$L^2(\RNumb^4,d^4x)$, where the fourth dimension is time, treated here on the
same footing as the positions in the 3D-space. The fundamental difference is
that the scalar procuct in $L^2(\RNumb^4,d^4x)$, which represents the
probability amplitudes, contains integration over time -- more discussion about
this case is in Sec.~\ref{sec:Schr}.

These mappings can always be written in terms of the so-called quantum
operations or their generalizations. The formalism of quantum operations was
invented around 1983 by Krauss \cite{Krauss}, who relied on the earlier
mathematical works of Choi \cite{Choi}.

The projection evolution operators at the evolution step $\tau_n$ are
formally defined as a family of transformations from the quantum state
space (density operators space)
$\mathcal{T}^{+}_{1}(\mathcal{K}(\tau_{n-1}))$ to the space
$\mathcal{T}^{+}(\mathcal{K}(\tau_{n}))$,
\begin{equation}
\label{eq:FOp1}
\FOp(\tau_n;\nu,\cdot): 
\mathcal{T}^{+}_{1}(\mathcal{K}(\tau_{n-1})) \rightarrow 
\mathcal{T}^{+}(\mathcal{K}(\tau_{n})),
\end{equation}  
where $\mathcal{T}^{+}(\mathcal{K}(\tau))$ is the space of finite trace,
positive and self-adjoint operators acting in the Hilbert space
$\mathcal{K}(\tau)$, $\nu\in {\cal Q}_{n}$, with
${\cal Q}_{n} \equiv{\cal Q}_{\tau_{n}}$ being a~family of sets of quantum
numbers defining potentially available final states for the evolution from
$\tau_{n-1}$ to $\tau_{n}$.

We denote by $\FOp(\tau_n;\nu,\rho)$ the result of the action of the operator
$\FOp(\tau_n;\nu,\cdot)$ on the density operator $\rho$, such that
$\FOp(\tau_n;\nu,\cdot)\rho = \FOp(\tau_n;\nu,\rho)$.  The notation
$\FOp(\tau_n;\nu,\rho)$ is in some cases more appropriate because, in general,
the changes principle does not constrain the evolution operators to be linear.

To use the generalized L\"uders projection postulate as the principle
for the evolution, the operators $\FOp(\tau_n;\nu,\rho)$ have to be
self-adjoint, non-negative, and with finite trace:
\begin{eqnarray}
  && \FOp(\tau;\nu,\rho)^\dagger = \FOp(\tau;\nu,\rho), \label{FOp21} \\
  && \FOp(\tau;\nu,\rho) \geq 0,                        \label{FOp22} \\
  && \sum_{\nu\in{\cal Q}_\tau}
     \Trace(\FOp(\tau;\nu,\rho)) < \infty,              \label{FOp23}
\end{eqnarray}
for every state $\rho$.  These three conditions allow $\FOp$ to
transform the density operator $\rho$ into another density operator, as
is shown in Eq.~(\ref{NewState}) below.

\begin{figure*}
  \centering
  \includegraphics[width=0.8\textwidth]{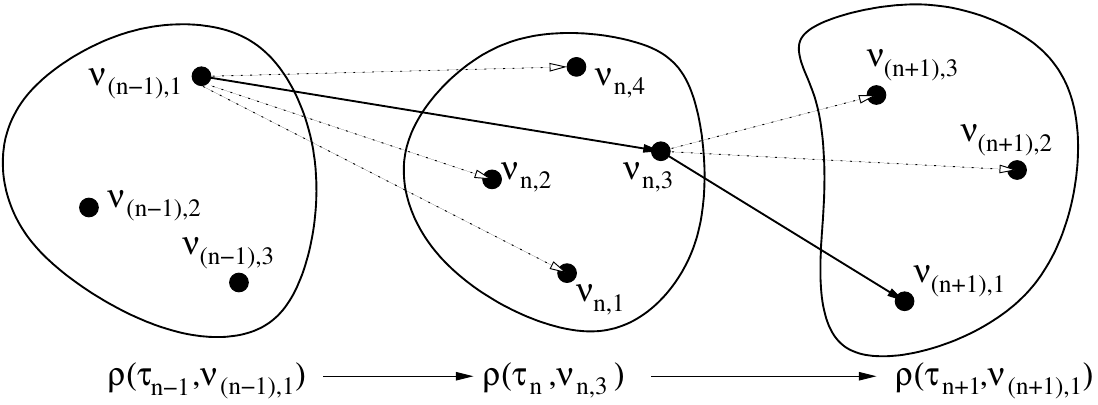}
  \caption{The density matrix $\rho$ is randomly chosen at each
    evolution step $\tau$ from the possible states labeled by
    $\mathcal{Q}_m = \{\nu_{m,1},\nu_{m,2},\dots\}$, where
    $m=n-1,n,n+1$.}
  \label{fig:rho}
\end{figure*}

Assume that at the evolution step $\tau_{n-1}$ the actual quantum state of
a~physical system is given by the density operator $\rho(\tau_{n-1};\nu_{n-1})$,
with $\nu_{n-1}\in{\cal Q}_{n-1}$. The changes principle implies that every step
of the evolution is similar to the measurement process in the sense that there
exists a~mechanism in the Universe, the chooser, which chooses randomly from the
set of states determined by the projection postulates the next state of the
system for $\tau=\tau_n$. With these assumptions, following Ref.~\cite{Lueders},
we postulate $\rho(\tau_n;\nu_n)$, $\nu_n\in{\cal Q}_n$, in the form \footnote{
  Remark: Eq.(\ref{NewState}) gives the set of the allowed states to which a
  physical system can randomly evolve from the state
  $\rho(\tau_{n-1};\nu_{n-1})$. }
\begin{equation}
\label{NewState}
\rho(\tau_n;\nu_n) =
            \frac{\FOp(\tau_{n};\nu_n,\rho(\tau_{n-1};\nu_{n-1}))}
{\textrm{Tr}\left(\FOp(\tau_{n};\nu_n,\rho(\tau_{n-1};\nu_{n-1}))\right)}.
\end{equation}
Because the chooser represents a~stochastic process, to fully describe
it one needs to determine the probability distribution for getting
a~given state in the next step of the evolution. An example of
an~evolution path is presented in Fig.~\ref{fig:rho} by the solid
line. The dotted lines show other potential paths.

%%%%%%%%%%%%%%%%%%%%%%%%%%%%%%%%%%%%%%%%%%%%%%%%%%%%%%%%%%%%
%%%%%%%%%%%%%%%%%%%%%%%%%%%%%%%%%%%%%%%%%%%%%%%%%%%%%%%%%%%%

In general, the probability distribution for the chooser is given by the
quantum mechanical transition probability from the previous to the next
state. This probability for pure quantum states is determined by the
appropriate probability amplitudes in the form of scalar products. The
transition probability among mixed states, in general, remains an open
problem, see e.g. \cite{Ulmann1976,Ulmann2016}.

We denote the transition probability (or the transition probability
density) from the state labelled by the set of quantum numbers
$\nu_{n-1}$ at $\tau_{n-1}$ to the state labelled by the set of quantum
numbers $\nu_n$ at $\tau_n$ for a given evolution process by
$\pev{\nu_{n-1} \to \nu_n}$. The arguments of $\pev{}$ indicate the
initial and the final state of the transition.

The most important realization of the evolution operators
$\FOp(\tau_n;\nu_n,\rho)$ can be constructed from the density matrix
$\rho$ and some operators $\EOp$ in the following form: for every
$\nu_n\in{\cal Q}_n$ we have
\begin{equation}
\FOp(\tau_n;\nu_n,\rho)=
\sum_\alpha
\EOp(\tau_n;\nu_n,\alpha) \ \rho \ \EOp(\tau_n;\nu_n,\alpha)^\dagger,
\label{EOperat}
\end{equation}
where the summation over $\alpha$ is dependent on the quantum numbers
$\nu_n$. It is easy to check that the conditions (\ref{FOp21}) and
(\ref{FOp22}) are automatically fulfilled, namely:
\begin{equation}
\FOp(\tau_n;\nu_n,\rho)^\dagger =
\sum_\alpha
\EOp(\tau_n;\nu_n,\alpha) \ \rho \ \EOp(\tau_n;\nu_n,\alpha)^\dagger
=\FOp(\tau_n;\nu_n,\rho) 
\label{F0p211} 
\end{equation}
and, since $\rho \geq 0$, we have for all $\phi \in \mathcal{K}$
\begin{eqnarray}
  && \Bra{\phi} \sum_\alpha
     \EOp(\tau_n;\nu_n,\alpha) \ \rho \ \EOp(\tau_n;\nu_n,\alpha)^\dagger
     \Ket{\phi}
     \nonumber \\
  && = \sum_\alpha \Bra{\phi} \EOp(\tau_n;\nu_n,\alpha) 
     \ \rho\ \EOp(\tau_n;\nu_n,\alpha)^\dagger \Ket{\phi} \geq 0.
\label{F0p221}
\end{eqnarray}
Using Eq.~(\ref{EOperat}) and the fact that trace is cyclic, the left
hand side of the condition (\ref{FOp23}) takes the form
\begin{eqnarray}
  && \sum_{\nu_n\in{\cal Q}_n} \sum_\alpha
     \Trace(\EOp(\tau_n;\nu_n,\alpha) \ \rho \
     \EOp(\tau_n;\nu_n,\alpha)^\dagger)
     \nonumber \\
  && = \sum_{\nu_n\in{\cal Q}_n} \Trace \left( \sum_\alpha
     \EOp(\tau_n;\nu_n,\alpha)^\dagger
     \EOp(\tau_n;\nu_n,\alpha) \ \rho \right) < \infty
\end{eqnarray}

Typical and useful examples of the $\EOp$ operators are connected with
the unitary evolution and the orthogonal resolution of unity. In the
first case the operator is
\begin{equation}
\label{ProjOpUnitary}
\EOp(\tau_n;\nu_{n0},\alpha_0)=U(\tau_n),
\end{equation}
where $\nu_{n0}$ and $\alpha_0$ are some fixed values of $\nu_n$ and
$\alpha$, and $U(\tau_n)$ is a~unitary operator. In this case, following
Eq.~(\ref{EOperat}), the next step of the evolution is chosen uniquely
with the probability equal to 1, as
\begin{equation}
\label{NewStateUnitary}
\rho(\tau_n;\nu_n) = 
U(\tau_n) \ \rho(\tau_{n-1};\nu_{n-1}) \ U(\tau_n)^\dagger.
\end{equation}
One needs to note that the unitary operator (\ref{ProjOpUnitary}) is not
parametrized by time but by the evolution parameter $\tau$, even though, in
general, it is time dependent \footnote{ The projection evolution is not a
  simple generalization of the traditional unitary evolution. One needs to
  remember, that the evolution parameter $\tau$ cannot be interpreted as time,
  it is only a parameter which enumerates quantum events.  The traditional form
  of the evolution, i.e. unitary evolution driven by time interpreted as a
  parameter is only an approximation which is valid if the following conditions
  are satisfied: a) average values of the time operator $\Aver{\hat{t}}_{n} =
  \Trace(\hat{t} \rho(\tau_n))$ are an increasing function of the evolution
  parameter, i.e., $\tau_{n-1} < \tau_{n}$ implies $\Aver{\hat{t}}_{n-1} <
  \Aver{\hat{t}}_{n}$; b) temporal spreads of subsequent states are very small,
  i.e., the variance $\Aver{(\hat{t}- \Aver{\hat{t}}_n)^2}_n \sim 0$; c) the
  probability of choosing the next state during the evolution is very close to
  1, i.e., $\pev{\nu_{n-1} \to \nu_n} \sim 1$, for all $n$ enumerating
  projection evolution steps of a system under consideration. } .

The PEv approach allows for the generalization of the idea of the unitary
evolution. For example, it is possible to consider the case when a few different
unitary evolution channels are opened, each with a given probability $p_m$. In
this case, the state for the evolution step $n$ is a linear combination of the
products of different unitary evolutions of the previous state,
\begin{equation}
\ronk = \sum_{m=1}^N p_m\, U_m(\tau_n) \ronl U_m(\tau_n)^\dagger \, ,
\end{equation} 
where $U_m^\dagger=U_m^{-1}$ for $m=1,\dots, N$.

In the case of the orthogonal resolution of unity with respect to the
quantum numbers $\nu_n$ the following conditions hold (we have fixed for
simplicity the $\alpha$ parameter and omitted it in the notation, but
the more general case can be written similarly):
\begin{eqnarray}
\label{ProjOpProjections}
&& \EOp(\tau_n;\nu_n)^\dagger=\EOp(\tau_n;\nu_n), \nonumber \\
&& \EOp(\tau_n;\nu_n)\EOp(\tau_n;\nu_n')
   = \delta_{\nu_n\nu_n'} \EOp(\tau_n;\nu_n), \nonumber  \\
&& \sum_{\nu_n \in {\cal Q}_{n}} \EOp(\tau_n;\nu_n)= \UnitOp,
\end{eqnarray}
where $\UnitOp$ denotes the unit operator. Different alternatives of choices of
the quantum states are described by different sets of quantum numbers $\nu_n$.

The probability distribution of choosing the next state of the evolution
generated by (\ref{ProjOpProjections}) is now given by the known quantum
mechanical formula:
\begin{equation}
\label{TransProbProjOp}
\pev{\nu_{n-1} \to \nu_n}=
\Trace\left( \EOp(\tau_n;\nu_n) \ \rho(\tau_{n-1},\nu_{n-1}) \ 
  \EOp(\tau_n;\nu_n)^\dagger \right).
\end{equation}
The above discussed examples, even though generic for many quantum
mechanical systems, are only special cases of the more general evolution
operators.

%%%%%%%%%%%%%%%%%%%%%%%%%%%%%%%%%%%%%%%%%%%%%%%%%%%%%%%%%%%%%%%%%%%%%%%%
\section{The quantum spacetime}
\label{rozdz.qspacetime}
%%%%%%%%%%%%%%%%%%%%%%%%%%%%%%%%%%%%%%%%%%%%%%%%%%%%%%%%%%%%%%%%%%%%%%%%

% \bibitem[Busch(1996)]{Busch1996}
% \bibitem[Gazeau(2015)]{Gazeau2015}

To simplifiy notation, we consider in this section a single evolution
step only, i.e., we keep the evolution parameter $\tau$ fixed. The quantum
spacetime, similarly to other properties of any quantum system can
change from one to another step of its evolution.

The projection evolution is compatible with any reasonable model of the
quantum spacetime. We consider here the four dimensional spacetime, but
the generalization to a different number of dimensions is
straightforward. In the following we do not consider relations between
quantum dynamics and geometrical description of the spacetime. It is a
very important problem which require further considerations and it is
postponed to future papers. In this paper, we apply a general PEv idea
only to the flat spacetime. Some simplified applications of this idea in
the non-flat spacetime by making use of the expectation values of
appropriate observables, instead of PEv evolution operators, can be
found in \cite{GPPSchwarz2022,GPPBKL2022-a,GPPBKL2022-b}.

In general, the full description of the Universe needs additonal
variables describing intrinsic properties of matter which, for
simplicity, we do not take into account, but they can be directly added
to the formalism.

Let $\StateSpace{K} \equiv \StateSpace{K}(\tau;\SpaceTime{X})$ denote the
Hilbert space, usually represented by square integrable functions on
$\SpaceTime{X}$ with respect to a given measure $\mu$. Let $\mathcal{A}$ be a
$\sigma$--algebra of $\mu$--measurable subsets of $\SpaceTime{X}$ so that
$(\SpaceTime{X},\mathcal{A}, \mu)$ represents a measurable space.  The set
$\SpaceTime{X}$ can be interpreted as a support of the classical spacetime.

%%%%%%%%%%%%%%%%%%%%%%%%%%%%%%%%%%%%%%%%%%%
\subsection{Generalized observables}

Actually the most general approach to quantum observables is given by
the formalism of positive operator vaued measures (POVM)
\cite{Busch1996}.  It is a~generalization of the orthogonal operator
valued measures equivalent to the use of the self-adjoint operators as
quantum observables. In our case, POVM allows to construct a common
measure of multidimensional obsevables, e.g., four vector operators.

In the following, we denote by $\mathcal{L}(\StateSpace{K})$ a set of
bounded operators on the state space $\StateSpace{K}$.

A positive, normalized to $\UnitOp$, operator valued measure (POV)
$\hat{M}:\mathcal{A} \to \mathcal{L}(\StateSpace{K})$ on
$(\SpaceTime{X},\mathcal{A})$ is defined as \cite{Busch1996}:
\begin{enumerate}
\item $\hat{M}(\Omega) \ge 0$ for all $\Omega \in \mathcal{A}$ (positivity);
\item if $\Omega_i \in \mathcal{A}$ is a countable set of disjoint sets then
$\hat{M}(\bigcup_i \Omega_i)= \sum_i \hat{M}(\Omega_i)$, the convergence is in
weak operator topology ($\sigma$--additivity);
\item $\hat{M}(\SpaceTime{X})=\UnitOp$ and $\hat{M}(\emptyset)=0$
(normalization).
\end{enumerate}
Such measures represent a contemporay notion of quantum observables. They
are related to physics by the so called ``minimal interpretation'' of quantum
mechanics \cite{Busch1996} which states that the expression
\begin{equation}
\label{eq:MinInterpQM}
\Prob(\hat{M}(\Omega);\hat{\rho})= \Trace(\hat{M}(\Omega) \hat{\rho})
\end{equation}
gives the probability that the observable $\hat{M}$ has a value in the
set $\Omega$ if the quantum system is in the state $\hat{\rho}$, where
$\hat{\rho}$ is a quantum density operator.

Let us now assume that in a given model we are able to define a POV
measure $\hat{M}_{ST}$ which describes positions in spacetime, i.e., the
operator $\hat{M}_{ST}(\Omega)$, where
$\mathcal{A} \ni \Omega \subset \SpaceTime{X}$, measures if the system
is in the spacetime region $\Omega$.

For a given observer $\mathcal{O}$ the spacetime $\SpaceTime{X}$ can be
decomposed into one dimensional time space $T$ and three dimensional
position space $S$, i.e., $\SpaceTime{X}=(T \times S)_{\mathcal{O}}$.
This decomposition allows to write the observable
$\hat{M}_T([x^{(0)}_A,x^{(0)}_B]) := \hat{M}_{ST}([x^{(0)}_A,x^{(0)}_B]
\times S)$ measuring if the quantum physical system is in the time
interval $[x^{(0)}_A,x^{(0)}_B]$, independently of its position in the
3D--space.
Such operator represents the time operator with respect to
the observer $\mathcal{O}$. In this context time is a component of a
compound observable representing a place in the spacetime. The
complementary operator $\hat{M}_S(Y) := \hat{M}_{ST}(T \times Y)$, where
$Y \subset S$, measures if the quantum system is in the region $Y$ of a
3D--space $S$, independently of its position on the time axis.

In this way, to every region $\Omega$ of the classical spacetime one can ascribe
the operator $\hat{M}(\Omega)$ measuring if the system is  in $\Omega$. 

Intuitively, a good observable could be also an operator $\hat{M}_{ST}(x)$
checking if the system is in a given point of the spacetime $x \in X$.  These
operators define a more natural correspondence between the classical spacetime
$\SpaceTime{X}$ and the quantum points $\hat{M}_{ST}(x)$.

Such operator can be imagined as a sequence of approximations
$\hat{M}_{ST}(\omega_k(x))$, where $\omega_k(x) \in \mathcal{A}$ is a
sequence of descending neighborhoods of the point $x$, i.e.,
$ \omega_1(x) \supset \omega_2(x) \supset \omega_3(x)\supset \dots
\omega_n(x)\supset \dots \supset \{x\}$. $\hat{M}_{ST}(\omega_k(x))$. It
is a very useful notion, one needs to remember, however, that the above
limit leads sometimes to the operator valued distributions. Despite that
one can construct also well behaving operators \cite{Gazeau2015}.

A natural connection between the POV measures $\hat{M}_{ST}(\Omega)$ and
$\hat{M}_{ST}(x)$ is given by
\begin{equation}
\label{eq:DensityPOV}
\hat{M}_{ST}(\Omega)
= \int_{\SpaceTime{X}} d\mu(x)\, \chi_{\Omega}(x)\, \hat{M}_{ST}(x) \,,
\end{equation}
where $\chi_{\Omega}(x)$ is the characteristic funcion of the set $\Omega$,
i.e., $\chi_{\Omega}(x)=1$ if $x \in \Omega$, otherwise $\chi_{\Omega}(x)=0$.

The last expression suggests that in many practical cases the components of the
position operator with respect to a given observer $\mathcal{O}$ for which
$x=(x^0,x^1,x^2,x^3)$ can be expressed as:
\begin{equation}
\label{eq:OperDensityPOV}
\hat{x}^{\mu}_{\mathcal{O}}
= \int_{\SpaceTime{X}_\mathcal{O} } d\mu(x)\,
x^\nu \, \hat{M}_{ST}(x) \, ,
\end{equation}
where $\SpaceTime{X}_\mathcal{O}= T \times S$ is the decompostion of the
spacetime into time and spatial part, with respect to the observer
$\mathcal{O}$.  The formula (\ref{eq:OperDensityPOV}) is compatible with
the integral quantization method, see \cite{Gazeau2015} and references
therein.

In general, the components of the spacetime position operator
(\ref{eq:OperDensityPOV}) do not commute and they do not have any common
eigenstates \cite{GPPSchwarz2022,GPPBKL2022-a,GPPBKL2022-b}. This
requires the construction of the preferred states representing the
quantum spacetime.

%%%%%%%%%%%%%%%%%%%%%%%%%%%%%%%%%%%%%%%%%%%%%%%%%%%%%%%%%%%%%%%%%%%%%%%
\subsection{Quantum spacetime points}

The construction of the preferred quantum spacetime states requires a
mapping $\kappa_{ST}$ between the support of the classical spacetime
$\SpaceTime{X}$ and $\StateSpace{K}$:
\begin{equation}
\label{eq:DefXK}
\kappa_{ST}: \SpaceTime{X} \ni x \to \Ket{\eta_{x}} \in \StateSpace{K} \, .
\end{equation}
One needs to remember that the state space consists of functions on the
spacetime support $\SpaceTime{X}$. However, the same set $\SpaceTime{X}$
serves as the set of labels indexing states in the mapping
(\ref{eq:DefXK}).

Using the coordinate frame corresponding to the observer $\mathcal{O}$,
Eq.~(\ref{eq:DefXK}) can be rewritten as
\begin{equation}
\label{eq:2DefXK}
\kappa_{ST_{\mathcal{O}}}:
\SpaceTime{X}_{\mathcal{O}} \ni (x^0,x^1,x^2,x^3)
\to
\Ket{\eta_{x^0,x^1,x^2,x^3}}_{\mathcal{O}} \in \StateSpace{K} \, ,
\end{equation}
where the coordinates $(x^0,x^1,x^2,x^3)$ of a point $x$ in spacetime
are quantum numbers enumerating the state representing this point with
respect to the observer $\mathcal{O}$. We call the vectors
$\Ket{\eta_{x}}= \Ket{\eta_{x^0,x^1,x^2,x^3}}_{\mathcal{O}}$, either the
position states or the quantum spacetime points.

This mapping has to fulfill two important conditions.

The main requirement, the selfconsistency of the position states, is to get the
appropriate expectation values of the position operators constructed with
respect to a given observer $\mathcal{O}$, i.e., we require to reproduce the
classical position values as the mean values of the position operators
$\hat{x}^\mu_{\mathcal{O}}$\,:
\begin{equation}
\label{eq:AverPosOper}
\Aver{\hat{x}^\mu_{\mathcal{O}};\eta_{x}} :=
\Bra{\eta_{x}} \hat{x}^\mu_{\mathcal{O}} \Ket{\eta_{x}} = x^\mu \, ,
\end{equation}
where $x=(x^0,x^1,x^2,x^3)$ with respect to $\mathcal{O}$ and the expectation
value of the operator $\hat{A}$ in the pure state $\psi$ is defiend as
\begin{equation}
\label{eq:ExpectVal}
\Aver{\hat{A};\psi}:= \Bra{\psi}\hat{A}\Ket{\psi} \, .
\end{equation}

The second condition comes from the observation that every physical object has
to be located somewhere in spacetime. This implies that the spacetime
position states have to furnish a resolution of unity:
\begin{equation}
\label{eq:XResUnity}
\int_\SpaceTime{X} d\mu(x) \Ket{\eta_{x}}\Bra{\eta_{x}}= \UnitOp \, .
\end{equation}
An important characteristics of the spacetime position states are variances of
the position observables $\Var(\hat{x}^\mu_{\mathcal{O}};\eta_{x})$, where the
variance of the operator $\hat{A}$ in the state $\psi$ is defined as
\begin{equation}
\label{eq:VarDef}
\Var(\hat{A};\psi)
:= \Aver{(\hat{A} - \Aver{\hat{A};\psi})^2;\psi} \,.
\end{equation}
The variances determine the ``sizes'' of the quantum points in
spacetime.

In the following, to simplify notation we fix a given observer
$\mathcal{O}$ and the index $\mathcal{O}$ will be ommited.

If the components of the spacetime position observable commute a
possible mapping can be defined by common eigenstates of the position
operators $\hat{x}=(\hat{x}^0,\hat{x}^1,\hat{x}^2,\hat{x}^3)$, where
$\hat{x}^{\mu} \Ket{\eta_{x^0,x^1,x^2,x^3}} = x^\mu
\Ket{\eta_{x^0,x^1,x^2,x^3}}$. The difficulty is that in some cases the
eigenstates of the spacetime position operators do not belong to the
Hilbert state space. In this case, usually, not all required expressions
are well defined, e.g., the expectation values of the position operators
within the Dirac delta type states are indetermined because the square
of the Dirac delta distribution does not exist.  Obviously, such
problems are already well recognized and can be solved by some
regularization procedures. From the physical point of view such space
usually constists of ortohogonal, i.e. independent, eigenstates with
extremely sharp localization. Such quantum states represent, in fact, a
structureless spacetime (the points are of ``size'' 0).

In the models where a set of states representing quantum points of a
spacetime is different from eigenstates of the position operators, the
product of variances (\ref{eq:OperDensityPOV}) is bounded from below by
the Heisenberg uncertainty principle\,:
\begin{equation}
\label{eq:HeisenberUncert}
\Var(\hat{x}^\mu;\eta_{x})  \Var(\hat{x}^\nu;\eta_{x})
\ge
\frac{1}{4}  \Aver{i[\hat{x}^\mu,\hat{x}^\nu];\eta_{x}}^2 \, .
\end{equation}
In such  models  we always deal with smeared quantum points which are not point
like objects. It is an important property, especially in the context of possible
singularities of the dynamics in spacetime.

%%%%%%%%%%%%%%%%%%%%%%%%%%%%%%%%%%%%%%%%%%%%%%%%%%%%%%%%%%%%%%%%%%%%%%%
\section{A quantum spacetime generated by a set of commuting mutiplication type
  position operators -- the quantum Minkowski spacetime}
\label{rozdz.minkowski}
%%%%%%%%%%%%%%%%%%%%%%%%%%%%%%%%%%%%%%%%%%%%%%%%%%%%%%%%%%%%%%%%%%

In this section we consider the simplest and at the same time the basic example
of spacetime.  The structureless quantum Minkowski spacetime is generated by a
set of the spacetime position operators in the state space
$\StateSpace{K}=L^2(\RNumb^4,d^{4}x)$, where $\RNumb^4$ is the support of this
spacetime. In this example we assume that these operators are commuting,
multiplication type operators which, with respect to a fixed but arbitrary
observer $\mathcal{O}$, can be written as
\begin{equation}
\hat{x}^\mu = \int_{\RNumb^4} d^4x\, x^\mu\, M_X(x) \, ,
\end{equation}
where $\Ket{\eta}_x:=\Ket{x}$ are generalized eigenstates of the traditional
positon operators $\hat{x}^\mu f(x^0,x^1,x^2,x^3)= x^\mu f(x^0,x^1,x^2,x^3)$ and
the operators $M_X(x)=\Ket{x}\Bra{x}$ give the resolution of unity of the
four-vector position operator
$\hat{x}=(\hat{x}^0,\hat{x}^1,\hat{x}^2,\hat{x}^3)$. Note, that the generalized
eigenstates $\Ket{x}$ are of the Dirac-delta type.

The observable $\hat{x}^0$ represents the quantum time and the remaing
operators $\hat{x}^k$, $k=1,2,3$ represent 3D-space position.

The operators
\begin{equation}
\hat{M}_{ST}(\Omega) = \int_{\RNumb^4} d^4x\, \chi_{\Omega}(x) M_X(x), 
\end{equation}
where $\Omega \subset \SpaceTime{X}$, give the orthogonal operator valued
measure which describes localization of points in the quantum Minkowski
spacetime.

The generated quantum states $\Ket{\eta}_x=\Ket{x}$ representing points of the
Minkowski space are orthogonal. There are no transitions among them and all the
dynamical structure has to be introduced in it from the outside.

Let us consider a~test particle in the Minkowki quantum spacetime. By
definition of the test particle we assume no back-reaction of the
particle onto the spacetime.

The scalar product in this state space $\StateSpace{K}$ is given by
\begin{equation}
\label{eq:ScalarProdR4}
\BraKet{\Phi_2}{\Phi_1}=\int_{\mathcal{K}} d^4x\, \Phi_2(x)^* \Phi_1(x).
\end{equation}
This scalar product is invariant with respect to the Lorentz transformations.

The scalar product (\ref{eq:ScalarProdR4}) has the following probabilistic
interpretation: the spacetime realization $\Psi(x)=\BraKet{x}{\Psi}$ of any pure
state $\Ket{\Psi} \in \StateSpace{K}$ represents the probability amplitude of
finding the particle in the spacetime point $x$, i.e., $|\Psi(x)|^2$ is the
probability density of finding this particle at $x$.

In general, the PEv approach leads to the breaking of the classical
causality. The functions
$\Psi(x) :=\BraKet{x^0,x^1,x^2,x^3}{\Psi} \in \mathcal{K}$, in their general
form, connect also events with space-like intervals $(x^0)^2-\vec{x}^2 <
0$. Obviously, this can be easily removed by assuming that $\StateSpace{K}$
consists of functions with time-like and zero-like support only, which means
that outside the set $(x^0)^2-\vec{x}^2 \ge 0$ the functions $\Psi(x)$ are
zero. Some experimental works \cite{epr-speed} suggest, however, that it is
a~natural phenomenon that the classical causality is broken in the quantum
world. To be more general, we allow for states which break the classical
causality to some acceptable extend. Within the PEv approach the quantum
causality is realized by keeping the correct sequence of the subsequent steps of
the evolution, ordered by the parameter $\tau$.

In general, the notion of simultaneity is observer dependent. However,
for every fixed choice of coordinates, in which one can distinguish
between space and time, one can construct a~spectral
measure $M_T(x^0)$, which for any fixed time $t=x^0$ projects onto the
space of simultaneous events:
\begin{equation}
  \label{NonRelSimultEvents}
  M_T(x^0) = \int_{\RNumb^3} d^3x \, M_X(x).
\end{equation}
This allows to interpret the operator $\hat{t} \equiv \hat{x}^0$ as the time
operator (for non-relativistic case a~preliminary attempt can be found in
Ref.~\cite{Debicki2007}) in the form
\begin{equation}
  \label{NonRelTimeOp}
  \hat{t} \equiv \hat x^0 = \int_{\RNumb} dx^0 \, x^0 M_T(x^0) \, .
\end{equation} 
which implies
\begin{equation}
  \label{eq:NonRelTOpMutipl}
  \hat{t}\Psi(x) \equiv \Bra{x}\hat{t}\Ket{\Psi}=
  \int_{\RNumb^4} d^4{x'} t' \BraKet{x}{x'}\BraKet{x'}{\Psi}= x^0 \Psi(x),    
\end{equation}
where the normalization of the position states $\Ket{x}$ is given by
$\BraKet{x}{x'}=\delta^4(x-x')$.

The spectral decompositions (\ref{NonRelSimultEvents}) and
(\ref{NonRelTimeOp}) allow to determine an ideal clock. However, the
more realistic clocks should be described by POV measures. A good
introduction to the discussion about clocks can be found in
Refs.~\cite{th-rel-mech-a,th-rel-mech-b,th-montevideo} and references
therein. We postpone this discussion to a future paper.

In the relativistic physics, the time operator is well determined only for a
given observer but it cannot be considered a~standalone observable, as it is
possible in the non-relativistic case. It always has to be treated as a~part of
the four-vector position operator $\hat{x}$.

As a by-product of the above considerations one can construct the spectral
measure which can be used as a measure of causality of a given state
$\Ket{\Psi}$ at the time $x^0$,
\begin{equation}
  \label{CausalTimeOperator}
  M_T^{(C)}(x^0) = \int_{C(x^0)} d^3x \,  M_X(x) \,, 
\end{equation}
where $C(x^0)=\{\vec{x}: (x^0)-\vec{x}^2 \ge 0 \}$. The expectation
value of this operator,
\begin{equation}
  \label{eq:CausalityMeasure}
 \mathrm{Prob}_C[\Psi]=\Bra{\Psi} M_T^{(C)}(x^0) \Ket{\Psi},
\end{equation}
gives the probability that the particle described by the state
$\Ket{\Psi}$ is in the light cone, both in the past and in the future
directions, with the vertex at $x^0$.

An important operator related to the time operator is the temporal component
$\hat{p}_0$ of the four-momentum operator
$\hat{p}=(\hat{p}_0,\hat{p}_1,\hat{p}_2,\hat{p}_3)$. In the spacetime
representation, the operator, which is canonically conjugate to the position
operator $\hat{x}^\mu$, is the generator of translations in the spacetime of a
single particle in the $\mu$ direction,
\begin{equation}
  \label{eq:FourMomOp}
  \hat{p}_\mu= i \frac{\partial}{\partial x^\mu}.
\end{equation}
To keep a consistent interpretation, the temporal component of the momentum
operator should measure, similarly to the spatial components, the value of the
product ``temporal inertia'' $\times$ ``speed in time'' for the particle moving
along the time direction.

In addition, because the temporal linear momentum is a component of the four
momentum operator, it determines the arrow of time: one direction corresponds to
$p_0>0$, the opposite direction to $p_0<0$.

The traditional interpretation of $p_0$ as the energy holds only in the case
when the equations of motion relate $p_0$ directly to the energy of the system,
like in the Schr\"odinger equation $\hat{p}_0=\Ham$, $\Ham$ being the
Hamiltonian. Similar relation is present in the relativistic Klein-Gordon
equation, $p_0^2= m_0^2 + \vec{p}^2$. This type of relations exists also for
other physical systems. In general, one can expect that in the spacetime
representation, the equation of motion of a free particle relates its
four-position to its four-momenta, with the possibility that also other degrees
of freedom, if present, can be involved.

Both the Schr\"odinger and the Klein-Gordon equations of motion allow to
indirectly measure the temporal component $p_0$ of the four-vector momentum
operator $\hat{p}$. It is traditionally expected that in our world the temporal
momentum $p_0 \ge 0$, even though this feature does not follow from the
mathematical structure of the model, as the $\hat{p}_0$ operator has the full
spectrum $\RNumb$.

The condition $p_0 \ge 0$ can be imposed either by assuming that the
equation of motion allows for real motion only if $p_0 \ge 0$, or that
this condition is a more fundamental property of our part of the
Universe. A~simple argument, or rather a hypothesis, supporting the
latter possibility is related to the initial state of our
Universe. Assuming that the four-momentum is a conserved quantity, the
initial chaotic motion of matter should have lead to the situation in
which matter moved in the $p_0>0$ and $p_0<0$ directions with the same
probability. The spatial components lead to the expansion of matter in
the $\RNumb^3$ space, the temporal component of the four-momentum,
however, lead to the separation of the Universe into two parts: one of
which is moving in the positive direction of time, while the other in
the negative direction of time. Both subspaces of states are orthogonal
and cannot communicate unless an interaction connecting both time
directions occurs. This implies that our part of the Universe
corresponds to one of the directions of the time flow, say, $p_0>0$. It
does not mean, obviously, that in our part of the Universe we do not
have the possibility to create particles with $p_0<0$. According to
common interpretation, such objects are antiparticles. This strongly
simplified picture requires further analysis but can provide a~possible
explanation of the $p_0>0$ phenomenon.

An interesting feature of the pair of the operators $\hat{x}$ and $\hat{p}$ is
that, since they fulfill the canonical commutation relations
\begin{equation}
  \label{eq:CommutRelXP}
  [\hat{p}_\mu, \hat{x}^\nu]= i\delta_\mu^\nu,
\end{equation}
they obey the Heisenberg uncertainty principle in the Robertson form
\cite{Robertson1929},
\begin{equation}
  \label{eq:PXUncertPrinc}
\Var(p_\mu;\psi) \ \Var(x^\nu;\psi)
  \ge \frac{1}{4} \Aver{i[\hat{p}_{\mu},\hat{x}^{\nu}];\psi}^2 
  = \frac{1}{4} \delta_\mu^\nu \, .    
\end{equation}
It is interesting to revisit in the future different forms of
the uncertainty principles for time, temporal component of the linear
momentum, and other observables.

An interesting example is the mass operator. Assume that the mass
operator for a~free particle is given by
\begin{equation}
  \label{eq:1MXUncertPrinc}
\hat{m}^2= \hat{p}_\mu \hat{p}^{\mu}.
\end{equation} 
Then, the uncertainty relation between the invariant mass and the
position in spacetime is given by
\begin{equation}
  \label{eq:2MXUncertPrinc}
\Var(m^2;\psi) \ \Var(x^\nu;\psi) \ge \Aver{p^\nu;\psi}^2.
\end{equation}
The width of such a~mass is bounded by the ratio of the expectation
value of $\Aver{p^\nu;\psi}$ and the variance $\Var(x^\nu;\psi)$.

In the case when $p_0$ is related to the energy by means of the
equations of motion for a given system, one obtains in a natural way the
uncertainty relation between the energy and time. For example, in the
case of the Schr\"odinger type of motion, described by the equation of
motion $\hat{p}_0 \Ket{\psi}=\Ham \Ket{\psi}$, the Heisenberg relation
(\ref{eq:PXUncertPrinc}) can be rewritten as
\begin{equation}
  \label{eq:HTUncertPrinc}
  \Var(\Ham;\psi) \ \Var(\hat{x}^0;\psi) \ge \frac{1}{4}.    
\end{equation}
This relation is fulfilled in the space of solutions $\Ket{\psi}$ of the
Schr\"odinger equation. Similar relations between time and energy can
always be obtained from appropriate equations of motion of the system
under consideration.

%%%%%%%%%%%%%%%%%%%%%%%%%%%%%%%%%%%%%%%%%%%%%%%%%%%%%%%%%%%%%%%%%%%%%%%%%
\section{Generators of the projection evolution}
%%%%%%%%%%%%%%%%%%%%%%%%%%%%%%%%%%%%%%%%%%%%%%%%%%%%%%%%%%%%%%%%%%%%%%%%

Within the traditional approach, the evolution of a~quantum state is
driven by a~Hamiltonian dependent operator $e^{-i{\cal \Ham}t}$. In the
projection evolution mechanism the changes of the system are spontaneous
and time is only an intrinsic variable of the physical system.

In this section we introduce a tool which facilitates the construction
of the evolution operators in terms of the projection operators. We
assume that a~subset of the evolution operators can be obtained from the
appropriate operators $\WOp$, the generators of the projection
evolution.

\textbf{For a~given evolution step $\tau$ the projection evolution generator
$\WOp(\tau)$ is defined as a self-adjoint operator which spectral
decomposition gives the orthogonal resolution of unity representing the
set of evolution operators} \footnote{
  Assuming discrete spectrum of an evolution generator $\WOp(\tau_n)$
  for a given evolution step the spectral theorem gives the following
  relation between $\WOp$ and the evolution operators $\EOp$:
  $\WOp(\tau_n)= \sum_\nu w_\nu \EOp(\tau_n) $. Note, that under rather
  weak conditions for a function $f$ the function of the generator
  $f(\WOp)= \sum_\nu f(w_\nu) \EOp(\tau_n)$ leads to the same evolution
  operators.  In the case of continuous spectrum one needs to use the integral
  form of the spectral theorem.
} .  % footnote

This generator can be subject to different constraints coming from
physics of the system under consideration. 

Let us consider a free single particle with spin equal to zero and no
intrinsic degrees of freedom. In this case the generator $\WOp$ can be
dependent on the spacetime position $\hat{x}$ and the four-momentum
$\hat{p}$ operators only.

Taking into account the translational symmetry in our Minkowsky spacetime, the
dependence of $\WOp$ on the position operators disappears. Imposing the
additional requirement of the rotational symmetry for this evolution
generator results in the construction of the operator $\WOp$ as a
function of the rotational invariants of the form
$a^\mu \hat{p}_\mu,\ a^{\mu\nu} \hat{p}_\mu \hat{p}_\nu, \dots$, where
$a^\mu,\ a^{\mu\nu},\dots$ are appropriate tensors with respect to the
$\Group{SO(3)}$ group. Basing on the experience of classical and quantum
physics one can expect that the expansion up to the second order in
momenta should be a~good approximation, which leaves us with
\begin{equation}
  \label{eq:WOpFreeParticle1}
  \WOp \EqCond{C} a^\mu \hat{p}_\mu + a^{\mu\nu} \hat{p}_\mu \hat{p}_\nu,
\end{equation}
where $\EqCond{C}$ means that $\WOp$ is equal to the right-hand side of
Eq.~(\ref{eq:WOpFreeParticle1}) only if the set of additional conditions
$C$ is fulfilled. This conditions depend on the physical properties of
the studied case. We will use that in Sec.~\ref{sec:symmetries} where
the symmetries are discussed.

The additional symmetries expected for a~free particle are the space
inversion and the antiunitary time reversal. Assuming that
$a^\mu,\ a^{\mu\nu},\dots$ are invariant with respect to both of these
symmetries, the linear term in momenta reduces to $a^0 \hat{p}_0$. The
quadratic term splits into two parts
$a^{00}(\hat{p}_0)^2 + a^{mn} \hat{p}_m \hat{p}_n $, where
$m,n=1,2,3$. The spatial quadratic term has no preferred direction
implying, that it can be written in the form $a^{mn}=B \delta^{mn}$,
which casts $\WOp$ in the form
\begin{equation}
  \label{eq:WOpFreeParticle2}
  \WOp \EqCond{C} a^0 \hat{p}_0 + a^{00} (\hat{p}_0)^2
  + B(\hat{p}_1^2+\hat{p}_2^2+\hat{p}_3^2).  
\end{equation}
To compare Eq.~(\ref{eq:WOpFreeParticle2}) with the standard quantum
mechanics, one can rescale it setting $a^0=1$. Then, the first and the
third term represent the Schr\"odinger equation for a~free particle with
mass $m=\frac{1}{2B}$. The second term is proportional to the second
time derivative $(p_0)^2 \sim -\frac{\partial^2}{\partial t^2}$ and is
not a~part of the Schr\"odinger equation in the standard formulation. It
is probably highly suppressed by the $a^{00}$ coefficient. By setting
this coefficient to zero we can remove this term from the equation,
recreating the standard Schr\"odinger evolution.

Similarly, imposing the Poincar\'e group invariance of $\WOp$, one has
to reject the first order term completely. Setting
$a^{\mu\nu}=g^{\mu\nu}=\mathrm{diag}(+1,-1,-1,-1)$ we are left with
\begin{equation}
  \label{eq:WOp-KG}
  \WOp_{KG} \EqCond{C} \hat{p}_\mu \hat{p}^\mu,
\end{equation}
which leads to the Klein-Gordon equation $\hat{p}_\mu \hat{p}^\mu=m^2$
with potentially additional conditions $C$. Assuming that $C$ stands for
positive mass $m>0$ and positive temporal component of the momentum
operator $p_0>0$, the generator (\ref{eq:WOp-KG}) describes the
evolution of a~free scalar particle. Changing the set of conditions $C$,
one can generate the evolution of other scalar objects. If
$a^\mu,\ a^{\mu\nu},\dots$ are some tensor operators, one can reproduce
other equations of motion. For example, in the case of
spin-$\frac{1}{2}$ particles, assuming $a^\mu=\gamma^\mu$, where
$\gamma^\mu$ are Dirac matrices, one gets the Dirac equation
\begin{equation}
  \WOp_D \EqCond{C} \gamma^\mu \hat{p}_\mu.
  \label{eq:WOp-D}
\end{equation}
We conclude that the known equations, which describe specific quantum
particles, are some special forms of the evolution operator $\WOp$,
which allows also to describe much more complicated cases.

%%%%%%%%%%%%%%%%%%%%%%%%%%%%%%%%%%%%%%%%%%%%%%%%%%%%%%%%%%%%%%%%%%%%%%%
\subsection{The Schr\"odinger evolution as a special case of PEv}
\label{sec:Schr}

The generator of the Schr\"odinger evolution can be written as
\begin{equation}
  \WOp_S = i \frac{\partial}{\partial t}-\Ham = \hat{p}_0 -\Ham.
  \label{gener:eq.2}
\end{equation}
Let us assume that the Hamiltonian $\Ham$ is independent of time. The
eigenvalues and the corresponding orthonormal eigenvectors of $\Ham$
will be denoted by $\epsilon_n$ and $\phi_{n\mu}(\vec{x})$,
respectively, such that
\begin{equation}
  \label{eq:HamDiscrSpectr}
  \Ham \phi_{n\mu}(\vec{x})=\epsilon_n \phi_{n\mu}(\vec{x}).
\end{equation}
The action of $\WOp_S$ on the full wave function results in
\begin{equation}
  \label{eq:WGenDiscrSpectr}
  \WOp_S\, \eta_{k_0}(x^0)\phi_{n\mu}(\vec{x})=
  w(k_0,n)\, \eta_{k_0}(x^0)\phi_{n\mu}(\vec{x}),
\end{equation}
where
\begin{eqnarray}
  && w(k_0,n)=k_0-\epsilon_n, \\
  && \eta_{k_0}(x^0)=\frac{1}{\sqrt{2\pi}} e^{-i k_0 x^0}.
\end{eqnarray}
The spectral decomposition of the generator $\WOp_S$ in the form of
a~Riemann-Stieltjes integral can be written as
\begin{equation}
  \WOp_S = \int_\RNumb w\, dE_{\WOp}(w),
  \label{gener:eq.3} 
\end{equation}
where $dE_{\WOp}(w)$ projects onto the eigenspace of $\WOp_S$ belonging
to the eigenvalue $w$. This subspace is spanned by the generalized
eigenfunctions of the form
\begin{equation}
  \label{eq:FunSpanSpaceW}
  \Phi_w(x^0,\vec{x})= \frac{1}{\sqrt{2\pi}}
  \sum_n \sum_\mu c_{n\mu} e^{-i (\epsilon_n+w) x^0}\phi_{n\mu}(\vec{x}), 
\end{equation}
with $c_{n\mu}$ being c-number coefficients. The scalar product in the
state space is given by (\ref{eq:ScalarProdR4}). Note that in the
traditional three-dimensional scalar product the integration over time
is absent,
\begin{equation}
  \label{eq:ScalarProdR3}
  \BraKet{\Phi_2}{\Phi_1}_3 =
  \int_{\RNumb^3} d^3x\, \Phi_2(x^0,\vec{x})^* \Phi_1(x^0,\vec{x}),
\end{equation}
because the state space $\StateSpace{K}_3=L^2(\RNumb^3)$ does not
contain time.

Using the scalar product (\ref{eq:ScalarProdR4}) we see that the
eigenfunctions (\ref{eq:FunSpanSpaceW}) are normalized to the Dirac delta
functions,
\begin{equation}
  \BraKet{\Phi_{w'}}{\Phi_{w}} = \int_{\RNumb^4} dx^0\, dx^1 dx^2 dx^3
  \Phi_{w'}(x^0,\vec{x})^* \Phi_{w}(x^0,\vec{x}) = \delta(w'-w).
  \label{ScalProdPhiS}
\end{equation}
There are a~few methods of obtaining vectors belonging to the state
space $\StateSpace{K}$. For example, one can consider the extended
Schr\"odinger equation which contains the temporal part describing the
temporal dependencies of the kinetic and potential terms. A possible,
but not the most general, such extension is given by the generator
\begin{equation}
  \label{eq:ExtSchrEq}
  \WOp_{GS}(\tau)=\hat{p}_0 - \Ham(\tau) +
  \left[\frac{1}{2} B^{-1}_T(\tau) \hat{p}_0^2 + V_T(\tau,x^0) \right],
\end{equation}
where, in agreement with the PEv approach, the temporal parts of the
kinetic and potential terms were added. They represent the kinematics
and the possible localization of a~physical object on the time axis. The
parameter $B^{-1}_T(\tau)$ represents a kind of temporal inertia of the
physical object.

The eigenfunctions (\ref{eq:FunSpanSpaceW}) considered within the
traditional state space $\StateSpace{K}_3$ are general solutions of the
Schr\"odinger equation, where the eigenvalue $w$ determines the zero
value of the energy represented by the Hamiltonian $\Ham$. It follows
from the fact that the eigenequation for $\WOp_S$, from
Eq.~(\ref{gener:eq.2}), can be written in the form
\begin{equation}
  i\frac{\partial}{\partial t} \phi_w = (\Ham+w) \phi_w,
\end{equation}
which means that the arbitrary eigenvalue $w$ shifts the energy
spectrum. Of course, the physics in $\StateSpace{K}_3$ is independent of
the chosen value of $w$.

We conclude that an important difference between the PEv approach and
the traditional formulation of quantum mechanics lies in the
interpretation of the wave functions $\Psi(x^0,\vec{x})$. In the PEv
formalism the function $|\Psi(x^0,\vec{x})|^2$, where
$\Psi(x^0,\vec{x}) \in \StateSpace{K}$, represents the joined probability
density of finding the particle in the four-dimensional spacetime point
$(x^0,\vec{x})$. In the traditional form of quantum mechanics with time
being a~parameter, the function $|\Psi(x^0,\vec{x})|^2$, where
$\Psi(x^0,\vec{x}) \in \StateSpace{K}_3$, represents the conditional probability
density of finding the particle in the three-dimensional space point
$\vec{x}$, assuming that the particle is localized at time $x^0$.

%%%%%%%%%%%%%%%%%%%%%%%%%%%%%%%%%%%%%%%%%%%%%%%%%%%%%%%%%%%%%%%%%%%%%%
\subsection{Relativistic equations of motion}

To see that the PEv approach allows to describe the relativistic
evolution equations in a~more natural way than the (1+3)-formalism, it
is sufficient to consider the Klein-Gordon equation of motion for a~free
scalar particle.

We are using the Minkowski space with the metric tensor
\begin{equation}
\label{eq:MinkMetrTens}
g^{00}=1, \quad g^{11}=g^{22}=\gamma^{33}=-1, \quad
\text{otherwise} \ g^{\mu\nu}=0 
\end{equation}
We assume that all four-vectors are presented by their contravariant components
as $a=(a^0,a^1,a^2,a^3)=(a^0,\vec{a})$. 

The generator of the appropriate evolution is given by
(\ref{eq:WOp-KG}). The mass operator $\hat{m}^2=\hat{p}_\mu\hat{p}^\mu$
has the following continuous spectrum and generalized eigenvectors:
\begin{equation}
  \label{eq:KGEigenequation}
  \hat{p}_\mu\hat{p}^\mu \eta_{k}(x)= w \eta_{k}(x),
\end{equation}
where $\eta_{k}(x)=\exp(-i k_\mu x^\mu)/(4\pi^2)$,
$k=(k^0,k^1,k^2,k^3)$, $p_\mu= k_\mu$, and $w \in
\RNumb$. Comparing both sides of Eq.~(\ref{eq:KGEigenequation}) one gets
the relation $k_\mu k^\mu=w$. For each $w$ this relation determines the
subspace $\StateSpace{K}_w$ invariant under the Poincar\'e group,
corresponding to states with definite $w$, i.e., they belong to the mass
shell. This subspace consists of all generalized eigenvectors of the
mass operator (\ref{eq:KGEigenequation}) belonging to this mass
shell. They are of the form
\begin{equation}
  \label{eq:SolutionKGEq}
  \Phi_w(x)= \int_{\RNumb^4} d^4k \,\delta^4(k_\mu k^\mu - w) c(k)\eta_{k}(x)
= \int_{\StateSpace{K}_w} d^4k \, c(k)\eta_{k}(x)
\, , 
\end{equation}
where $c(k)$ is a function representing the profile of the wave package
(\ref{eq:SolutionKGEq}). 

This implies that the evolution operators generated by (\ref{eq:WOp-KG})
are the generalized projection operators
\begin{equation}
  \label{eq:KGEvolOp}
\EOp(\tau;w) = \int_{\StateSpace{K}_w} d^4k \Ket{\eta_{k}} \Bra{\eta_{k}}
\, .
\end{equation}
Using the usual conditions that the space of states is restricted to the
states for which $\hat{m}^2>0$ and $\hat{p}_0>0$, the eigenvalues $w$
are traditonally interpreted as the invariant mass squared, $m^2$.  In
this case, the evolution operators (\ref{eq:KGEvolOp}) can be rewritten as
\begin{equation}
  \label{eq:KGEvolOpPhys}
\EOp_C(\tau;w) = \int_{\StateSpace{K}_{w+}} d^4k 
\Ket{\eta_{k}}  \Bra{\eta_{k}},
\end{equation}
where $\StateSpace{K}_{w+}$ consists of functions $ \phi(k)
\in\StateSpace{K}_{w}$ with the constraints: $k^2 > 0$ and $k_0>0$.

In this case the evolution operators (according to the evolution
principle (\ref{NewState}) and the definition (\ref{EOperat}))
projecting on the subspaces $\StateSpace{K}_{w+}$ reproduce the known
solutions for the standard scalar particle of non-zero mass,
\begin{equation}
  \label{eq:SolutionKGEqMass}
  \Phi_w(x)= \int_{\RNumb^3} \frac{d^3k}{k_0}
  c(\vec{k}) \eta_{(k_0,\vec{k})}(x),
\end{equation}
where $k=(k_0,\vec{k})=(\sqrt{m^2+\vec{k}^2},\vec{k})$. Note that both
vectors (\ref{eq:SolutionKGEq}) and (\ref{eq:SolutionKGEqMass}) are
normalized Dirac delta type distributions.

One can extend the evolution generator of the Klein-Gordon equation to the case
of a particle in the electromagnetic field by including the appropriate
four-vector field $A_\mu$. Using the minimal coupling scheme one gets
\begin{equation}
  \label{eq:ExtKGEqVecField}
  \WOp=(\hat{p}_\mu -A_\mu)(\hat{p}^\mu -A^\mu).
\end{equation}
This vector field can play a~role similar to the temporal part of the potential
in the extended Schr\"odinger PEv generator (\ref{eq:ExtSchrEq}) allowing, in
some cases, for solutions in the form of square integrable states.

All other relativistic equations of motion can be reproduced in
a~similar way. One needs, however, to remember that physical
consequences of the PEv approach are tremendous. First of all, time
becomes a~quantum observable and it has to be treated on the same
footing as the remaining position coordinates.

%%%%%%%%%%%%%%%%%%%%%%%%%%%%%%%%%%%%%%%%%%%%%%%%%%%%%%%%%%%%%%%%%%%%%%%%%%%%%%%%
\section{Symmetries}
\label{sec:symmetries}
%%%%%%%%%%%%%%%%%%%%%%%%%%%%%%%%%%%%%%%%%%%%%%%%%%%%%%%%%%%%%%%%%%%%%%%%%%%

As it is well known, different kinds of symmetries play a fundamental
role in physics. They are the most important constraints for structure,
interactions and motion of physical objects.

In the case of the PEv formalism one thinks about two distinct types of
symmetries:
\begin{enumerate}
\item[(A)] the symmetries for a~fixed step of the evolution, i.e., for
  a~constant evolution parameter $\tau$; 
\item[(B)] the symmetries related to the transition of the system from
  one step of the evolution to another, i.e., for the case when the
  evolution parameter changes, $\tau_{n-1} \to \tau_n$.
\end{enumerate}

The first type of symmetries (A) describes structural, spacetime and
intrinsic properties of a quantum system. An important difference is
that time is now a quantum observable. Taking this into account,
symmetry analysis seems to be similar to those performed in relativistic
quantum mechanics. Many results remain valid, but most of them require
reinterpretation.

The second type of symmetries (B) is different because the evolution
operators are involved in the symmetry analysis. The operators
$\FOp(\tau;\nu,\rho)$ can have different structures, they can be unitary
operators, projection operators and other type of operators which allow
to transform quantum states into new quantum states. This opens many
mathematical and interpretational problems.

In this section we analyze only two elementary properties related to symmetries
of the type (B) for the case of the fixed state space
$\StateSpace{K}=\StateSpace{K(\tau_n)}$, i.e., the state space is the same for
all $\tau_n$. More extended analysis is beyond the scope of this paper.

We consider the evolution operators $\FOp(\tau_{n};\nu_{n})$ for which
the operators $\EOp(\tau_{n};\nu_{n})$ form either an orthogonal
resolution of unity (\ref{ProjOpProjections}) or the generalized unitary
PEv operators (\ref{NewStateUnitary}). The other cases will be
considered in a subsequent paper.

The problem is to find these physical properties which remain invariant at
subsequent steps of the evolution.  In other words, we are looking for the
transformations from one evolution step to another, which do not change our
physical system.

We start by writing the definition of transformations of the operator
$\FOp(\tau_{n};\nu_{n},\ro{n-1})$ under an action of the group $\Group{G}$.  Let
us denote by $S: \Group{G} \to \StateSpace{K}$ a unitary operator
representation of the group $\Group{G}$ in the state space $\StateSpace{K}$.
The transformation of the evolution operator $\FOp$ is defined as:
\begin{equation}
\label{eq:FOpTrans0}
\FOp'(\tau_n;\nu_n,\roprime{n-1})=
S(g) \ \FOp(\tau_n;\nu_n,\ro{n-1}) \ S(g^{-1}) \, ,
\end{equation} 
where
\begin{equation}
\label{eq:FOpTrans1}
\roprime{n-1}=S(g) \ro{n-1} S(g^{-1}) \, .
\end{equation} 
This definition follows an idea of transformations of functions of more complex
objects, e.g., the rotation $\hat{\mathcal{R}}$ of a vector function
$f:\RNumb^3\rightarrow \RNumb^3$. The values of the rotated function $f'$ having
the rotated argument $x'$ should be equal to the rotation of the value of the
original function having the original argument, $f'(x')=\hat{\mathcal{R}}f(x)$.

The definition (\ref{eq:FOpTrans0}) can be expressed in a~more convenient form:
\begin{equation}
\label{eq:FOpTrans}
\FOp'(\tau_n;\nu_n,\ro{n-1})=
S(g) \ \FOp(\tau_n;\nu_n,S(g^{-1})\ro{n-1} S(g)) \ S(g^{-1}) \, .
\end{equation} 
A fundamental property of the definition (\ref{eq:FOpTrans0}) is that it allows
to conserve probability structure of transitions under action of the group
$\Group{G}$.  To show this feature, let us assume that the transition
probability from the evolution step $\tau_{n-1}$ to $\tau_{n}$ is given by
\begin{eqnarray}
\label{eq:1TransProb}
&&\pev{\ro{n-1} \rightarrow  \ro{n}=
  \frac{\FOp(\tau_n;\nu_{n},\ro{n-1})}{\Trace[\FOp(\tau_n;\nu_{n},\ro{n-1})]}}
\nonumber \\
&& = \Trace[\FOp(\tau_n;\nu_{n},\ro{n-1})] \, ,
\end{eqnarray}
as it is for the projection evolution operators represented by an orthogonal
resolution of the unit operator. Here, $\nu_n$ denotes the set of allowed
quantum numbers. Similarly, the transition probability from the state
$\roprime{n-1}=S(g) \ro{n-1} S(g)^{-1}$ to the state
$ \tilde{\rho}(\tau_{n},\nu_{n})$ obtained by the transformed evolution operator
$\FOp'$ is given by
\begin{eqnarray}
\label{eq:2TransProb}
&& \pev{\roprime{n-1} \rightarrow \tilde{\rho}(\tau_{n},\nu_{n})=
\frac{\FOp'(\tau_n;\nu_{n} \, ,\roprime{n-1})}{
  \Trace[\FOp'(\tau_n;\nu_{n} \, ,\roprime{n-1})]}}
\nonumber \\
&& = \Trace[\FOp'(\tau_n;\nu_{n} \, ,\roprime{n-1})] \, .
\end{eqnarray}
Because $\Trace(AB)=\Trace(BA)$, it follows from the equation
(\ref{eq:FOpTrans0}) that both probabilities are equal:
\begin{equation}
\pev{\ro{n-1} \rightarrow \ro{n}}
=
\pev{\roprime{n-1} \rightarrow \tilde{\rho}(\tau_{n},\nu_{n})} \, .
\end{equation}
A consequence of such symmetry for every step of the evolution is the fact that
the action of the group $\Group{G}$ does not change the probability structure of
the possible evolution paths. 

The second important problem is a relation between symmetries and
conservation laws.  Intuitively one can say that we are looking for
conditions under which the expectation value of a given observable $A$
is conserved during the projection evolution process: 
\begin{equation}
\Aver{A;\ro{1}}=\Aver{A;\ro{2}}=\dots=\Aver{A;\ro{n}}=\dots \, .
\end{equation}
The required conditions may involve special relations between the
evolution operators, density operators and quantum observables.

In the case when the evolution is described by the operators
$\EOp(\tau_n;\nu_{n},\alpha)$  the conservation of the expectation value
$\Aver{A;\ro{k}}$ has the following form
\begin{eqnarray}
\label{eq:ConsA}
&& \Trace[A\;\ro{n-1}] = \Big[ \Trace[A\;\ro{n}] \nonumber \\
&& = \frac{\sum_{\alpha} \Trace[A 
\EOp(\tau_n;\nu_{n},\alpha) \ \ro{n-1} \ \EOp(\tau_n;\nu_{n},\alpha)^\dagger]}{
\sum_{\alpha} \Trace[
\EOp(\tau_n;\nu_{n},\alpha) \ \ro{n-1} \ \EOp(\tau_n;\nu_{n},\alpha)^\dagger]}
\Big]
\, .
\end{eqnarray}
In the case when we have the unitary type operators \\
$\EOp(\tau_n;\nu_{n},\alpha) = \sqrt{p(\tau_n;\nu_n,\alpha)}
U(\tau_n;\nu_n,\alpha)$, where $p(\tau_n;\nu_n,\alpha) \ge 0$, the condition
(\ref{eq:ConsA}) can be rewritten as
\begin{eqnarray}
\label{eq:ConsUnitaryAver}
&& \Trace\left[A \ro{n-1} \right] 
= \frac{1}{\sum_{\alpha} p(\tau_n;\nu_n,\alpha)} \nonumber \\
&& \sum_{\alpha} p(\tau_n;\nu_n,\alpha)
\Trace[U(\tau_n;\nu_n,\alpha)^\dagger\ A\ U(\tau_n;\nu_n,\alpha)\ \ro{n-1}].
\end{eqnarray}
The expectation value of the observable $A$ is conserved if the operator $A$
commutes with the evolution operators, i.e.,
$[A,U(\tau_n;\nu_n,\alpha)]=0$. This fact has its counterpart in the standard
quantum mechanics -- if the Hamiltonian commutes with the operator $A$, the
expectation value $\Aver{A}$ is conserved during the unitary evolution generated
by this Hamiltonian. For the general case when $\EOp(\tau_n;\nu_{n},\alpha)$ is
an orthogonal decomposition of unity the condition (\ref{eq:ConsA}) to be
fulfilled requires more complicated relations between the evolution operators,
states and the observable $A$. This will be considered elswere. However, a
special case when the evolution operator and the quantum observable, are
invariant under a~given symmetry can be solved generally. For simplicity of
notation in the following we fix the index $\alpha$.

Let a compact Lie group $\Group{G}$ be a symmetry group of the evolution
generator $\WOp(\tau_n)$, i.e.,
$S(g)\WOp(\tau_n) S(g)^{-1}=\WOp(\tau_n)$ for every $g \in \Group{G}$
and every $\tau_n$, where the operators $S(g)$ play the role of
a~unitary operator representation of this symmetry group in the state
space $\StateSpace{K}$. Because the group $\Group{G}$ is the symmetry
group of the generator $\WOp(\tau_n)$, its eigenstates
$\Ket{\tau_n;\kappa_n \Gamma_n a}$ form the irreducible subspaces of the
irreducible representations of this group,
\begin{equation}
\label{eq:1SymW}
\WOp(\tau_n)  \Ket{\tau_n;\kappa_n \Gamma_n a_n}= 
w(\tau_n;\kappa_n \Gamma_n)  \Ket{\tau_n;\kappa_n \Gamma_n a_n},
\end{equation}
\begin{equation}
\label{eq:2SymW}
S(g)\Ket{\tau_n;\kappa_n  \Gamma_n a_n}
=  \sum_{a'_n} \Delta^{\Gamma_n}_{a'_n a_n}(g)
\Ket{\tau_n;\kappa_n \Gamma_n a'_n} \, ,
\end{equation} 
where $\Delta^{\Gamma_n}$ denotes the irreducible representation of the
symmetry group $\Group{G}$ labelled by $\Gamma_n$, the quantum number
$a_n$ labels vectors within the given irreducible representation
$\Delta^{\Gamma_n}$, the set of quantum numbers $\kappa_n$ describes
these properties of our quantum system which are independent of the
symmetry, and at the same time it distinguishes among equivalent
irreducible representations for fixed $\Gamma_n$. For every $\tau_n$ the
vectors $\Ket{\tau_n;\kappa_n \Gamma_n a_n}$ form the ortonormal bases
in the state space $\StateSpace{K}$ and the quantum numbers $\Gamma_n$
belong to an established set of labels
$\{\Gamma^{(1)},\Gamma^{(2)},\dots \Gamma^{(k)},\dots\}$ enumerating
irreducible representations of the group $\Group{G}$. This independent
of the evolution step set of labels can be determined from the
decomposition of the state space $\StateSpace{K}$ into irreducible
subspaces.

In this case the spectral decomposition of the evolution generator can be
written as
\begin{equation}
\label{eq:SpectrDecomWSym}
\WOp(\tau_n) = \sum_{\kappa_n,\Gamma_n}
w(\tau_n;\kappa_n,\Gamma_n) P(\tau_n;\kappa_n \Gamma_n) \, ,
\end{equation}
where $w(\tau_n;\kappa_n,\Gamma_n)$ are eigenvalues and the projectors on the
eigenspaces read
\begin{equation}
\label{eq:3SymW}
P(\tau_n;\kappa_n \Gamma_n) =  \sum_{a_n}
\Ket{\tau_n;\kappa_n \Gamma_n a_n}\Bra{\tau_n;\kappa_n \Gamma_n a_n} \, .
\end{equation} 
The decomposition (\ref{eq:SpectrDecomWSym}) determines the following
evolution operators
\begin{equation}
\label{eq:4SymW}
\EOp(\tau_n;\kappa_n \Gamma_n)  = P(\tau_n;\kappa_n \Gamma_n) \, 
\end{equation} 
for which the two orthogonality relations hold
\begin{equation}
\label{eq:4aSymW}
\EOp(\tau_n;\kappa_n \Gamma_n) \, \EOp(\tau_n;\kappa'_n \Gamma'_n) 
= \delta_{\Gamma_n,\Gamma'_n}\delta_{\kappa_n,\kappa'_n} \,
\EOp(\tau_n;\kappa_n \Gamma_n)  \, .
\end{equation} 
and 
\begin{equation}
\label{eq:4bSymW}
\EOp(\tau_{n};\kappa_n \Gamma_n) \EOp(\tau_{n'};\kappa_{n'} \Gamma_{n'}) 
= 0 \ \text{ if } \ \Gamma_n \neq \Gamma_{n'}
\end{equation} 
Using the above conditions, the Casimir operator $\mathcal{C}^2$ of the
symmetry group $\Group{G}$, which is an observable invariant with
respect to this symmetry group, satisfies
\begin{equation}
\label{eq:5SymW}
 \mathcal{C}^2 \EOp(\tau_n;\kappa_n \Gamma_n)
= c_{\Gamma_n} \EOp(\tau_n;\kappa_n \Gamma_n) \, ,
\end{equation}
where $c_{\Gamma_n}$ are eigenvalues of the Casimir operator obtained
from
\begin{equation}
\mathcal{C}^2 \Ket{\tau_n;\kappa_n  \Gamma_n a_n}
= c_{\Gamma_n}\Ket{\tau_n;\kappa_n  \Gamma_n a_n} \, .
\end{equation}
Let $\rho_0$ denote the initial state. After the first step of the
projection evolution one gets a new state
\begin{equation}
\label{eq:EvolFirstStep}
\rho(\tau_1; \kappa_1,\Gamma_1)
= \frac{
\EOp(\tau_1;\kappa_1\Gamma_1 ) \rho_0 
\EOp(\tau_1;\kappa_1\Gamma_1)}{
\Trace\left[\EOp(\tau_1;\kappa_1 \Gamma_1) \rho_0 
\EOp(\tau_1;\kappa_1 \Gamma_1) \right]}  \, .
\end{equation}
The expectation value of the Casimir operator is
\begin{equation}
\label{eq:2EvolFirstStep}
\Trace[\mathcal{C}^2 \rho(\tau_1; \Gamma_1, \kappa_1)] =
\frac{\Trace \left[
\mathcal{C}^2 \EOp(\tau_1;\kappa_1 \Gamma_1) \rho_0 
\EOp(\tau_1;\kappa_1\Gamma_1 ) \right] }{
\Trace\left[\EOp(\tau_1;\kappa_1 \Gamma_1) \rho_0 
\EOp(\tau_1;\kappa_1 \Gamma_1) \right]} = c_{\Gamma_1} \, .
\end{equation}
Because of the othogonality relations (\ref{eq:4aSymW}) and (\ref{eq:4bSymW})
one gets
\begin{equation}
\label{eq:2EvolSecondStep}
\Trace[\mathcal{C}^2 \rho(\tau_2; \Gamma_2, \kappa_2)] =
\frac{\Trace \left[
\mathcal{C}^2 \EOp(\tau_2;\kappa_2 \Gamma_2)
\rho(\tau_1; \kappa_1,\Gamma_1) 
\EOp(\tau_2;\Gamma_2 \kappa_2) \right] }{
\Trace\left[\EOp(\tau_2;\kappa_2\Gamma_2)
\rho_0 
\EOp(\tau_2;\kappa_2 \Gamma_2 ) \right]}
= \delta_{\Gamma_1,\Gamma_2} \, c_{\Gamma_1} \, .
\end{equation}
This implies that the value of this Casimir operator is fixed for all subsequent
steps
\begin{equation}
\Trace[\mathcal{C}^2 \rho(\tau_n; \Gamma_n, \kappa_n)] 
 = 
\begin{cases}
 c_{\Gamma_1}, & \text{for } \Gamma_n=\Gamma_1 \, , \\
        0,     & \text{for } \Gamma_n \neq \Gamma_1 \, .
\end{cases} 
\end{equation}
We conclude that if the evolution operators are invariant with respect
to the group $\Group{G}$ and fulfill the above conditions, the value of
the Casimir operator $\mathcal{C}^2$ of the group $\Group{G}$ is
conserved during the evolution.

This special case has its analogy in the standard quantum mechanics. Let
us assume that the Hamiltonian $\Ham$ is invariant with respect to a
group $\Group{G}$. The eigenvectors of $\Ham$ belong to the invariant
subspaces spanned by the bases of the irreducible representations of the
group $\Group{G}$. In this case the expectation value of the Casimir
operator is conserved during the unitary evolution generated by this
Hamiltonian.

We have presented a~short outline of some open problems related to the
symmetry analysis within the projection evolution approach. PEv opens
new areas for applications of symmetries and group theoretical methods
in physics.

%%%%%%%%%%%%%%%%%%%%%%%%%%%%%%%%%%%%%%%%%%%%%%%%%%%%%%%%%%%%%%%%%%%%%%%
\section{Concluding remarks}
%%%%%%%%%%%%%%%%%%%%%%%%%%%%%%%%%%%%%%%%%%%%%%%%%%%%%%%%%%%%%%%%%%%%%%%

The discussion about the structure and the role of time is as long as
the history of physics. A~collection of papers devoted to different
aspects of the physical time from the modern perspective can be found,
among others, in \cite{th-2slits-b,muga2}. In Ref.~\cite{th-2slits-b}
the paper by P.~Busch mentiones three types of time. The most popular
one is time considered as a parameter which is measured by an external
laboratory clock, uncoupled from the measured system. This time is
called the external time. Time can be defined also through the dynamics
of the observed quantum systems, in which case we deal with the
dynamical (or intrinsic) time. Lastly, time can be considered on the
same footing as other quantum observables, especially as positions in
space. This is called by P.~Busch the observable (or event) time and it
represents the approach considered in the present paper in which we
discuss the most natural model of quantum spacetime in which time is a
quantum observable and it is treated on the same footing as the 3D-space
position observables. Time considered here is an essential component of
the position in spacetime.  It is also important that it allows to
calculate temporal characteristics of a quantum system on the same basis
as it can be done for other observables.

In the experimental practice the external time is usually used. It is
introduced by constructing different kinds of semi-macroscopic clocks
(because of the required interface with the macroscopic world). They are
constructed in such a way to be uncoupled from the analyzed physical
phenomenon. Because in PEv approach the state of the clock at the
evolution step $\tau_n$ is, in principle, described by its density
operator $\rho(\tau_n;\nu)$, this type of clocks seems to keep the
ordering relation in the set of all values of the evolution parameter
$\tau$, i.e., it has to fulfil the relation
$\Trace{(\hat{t}\rho(\tau_{n+1};\nu'))} >
\Trace{(\hat{t}\rho(\tau_n;\nu))}$, where $\hat{t}$ is the time
operator. The trace $\Trace{(\hat{t}\rho(\tau_n;\nu))}$ denotes in
analogy to the average position of an object in the 3D-space the
expectation value of the temporal position, i.e., time measured by the
clock being in the state $\rho(\tau_n;\nu)$ at the step evolution
$\tau_n$.  Having one clock, one can treat it as the standard clock. All
other clocks can be constructed and synchronized to this standard
clock. In this context the external time, even though very useful, is
a~conventional rather than physical entity.

The intrinsic time, or times, to be more precise, is determined by a set
of appropriate dynamical variables. It is compatible with our ``changes
principle'', i.e., that changes of states or observables are more
fundamental than the time itself. However, because in our approach the
physical time is a quantum observable, the required characteristic times
(intrinsic times) for a given physical process can be directly
calculated. In this context, the intrinsic times are not fundamental but
derivable temporal observables.

In the PEv approach the spacetime is ``created'' in the same way as the
other quantum properties of our Universe. The positions in spacetime are
related to the eigenstates of the spacetime position measure. This
implies that the PEv idea leads to a~background free theory.  Such
approach is important not only for particle physics but also for
clasical and quantum relativity, and finally for unification of quantum
mechanics with gravity.

Restricting the discussion to the flat spacetime, we proposed
a~self-adjoint spacetime position operator which transforms as any
four-vector with respect to the Poincar\'e group. As a consequence of
the spectral theorem this operator defines the covariant and orthogonal
resolution of unity which determines the ideal spacetime position
measure for quantum events. Obviously, real physical devices
representing such measure cannot be ideal, and in practice this measure
has to be replaced by POVM type operators. An interesting discussion,
especially about clocks can be found in
\cite{th-rel-mech-a,th-rel-mech-b,th-montevideo} and references
therein. A part of this discussion should be revisited in the context of
PEv -- it is a subject for further studies. A related problem which
requires further analysis are spacetime frames which are natural
constructions in the PEv model, and which support a notion of general
covariance up to the transformations allowed among the quantum
observables
\cite{Cov-Hohn2018,Cov-Hohn2019,Cov-Giacomini2019,Cov-Vanrietvelde2018,Cov-Vanrietvelde2020,Nonloc-CastroRuiz2020}.

Spacetime represented in the PEv model leads to many important quantum
effects such as time interference. This interference seems to manifest
itself for times shorter than femtoseconds \cite{Gozdz2015}. The more
important condition is the relation between the time spread of the wave
function and the temporal distance between the openings of the
slits. The former should be larger than the latter.

More generally, this structure can, in a natural way, account for many
quantum mechanical effects like the delayed-choice experiments (for
Wheeler's paradox see Refs.~\cite{DealyedChoice2008,DelayedChoice2018}).

The time operator and the corresponding conjugate temporal momentum
operator are the very natural complements of the covariant relativistic
four-position and four-momentum operators. The temporal component
$\hat{p}_0$ of the momentum operator is also responsible for the basic
arrow of time represented by the operator ${\hat{p}_0}/{|\hat{p}_0|}$,
i.e, the sign of the temporal momentum determines the direction along
the time axis.

The corresponding components of the spacetime position operator and the
four-momentum operator fulfil the canonical commutation relations and as
a consequence they obey the standard Heisenberg uncertainty
relations. In addition, it introduces through the equations of motion
(which usually involve the temporal momentum) the time-energy
uncertainty relation.

One needs to notice that the observable time allows also, in a very
natural way, for the dependence of interactions on the temporal distance
among particles or quantum events. A schematic example of a potential of
this type is shown in Ref.~\cite{Gozdz2014}, but this problem is still
open.

The evolution generators presented in this paper are effective tools
which link different kinds of traditional equations of motions and the
PEv approach. They allow to construct the evolution operators
corresponding to the Schr\"odinger, Klein-Gordon, Dirac and other
equations of motions. However, one needs to remember about another
interpretation of quantum states in PEv with respect to time.

The idea of ``changes principle'' and the concept of quantum evolution
as a stochastic process driven by the evolution ordering parameter
$\tau$, not time, is much more general than the specific model presented
in this paper.  However, the proposed implementation seems to be the
minimal interpretation satisfying main requirements about the quantum
spacetime structure.

%%%%%%%%%%%%%%%%%%%%%%%%%%%%%%%%%%%%%%%%%%%%%%%%%%%%%%%%%%%%%%%%%%%%%%%
%\bibliographystyle{plain}
\bibliographystyle{ieeetr}
\bibliography{PEv}

\begin{thebibliography}{10}

\bibitem{pauli-a}
W.~Pauli, ``Quantentheorie,'' in {\em Quanten, Handbuch der Physik} (H.~Geiger
  and K.~Scheel, eds.), pp.~1--278, Berlin, Heidelberg: Springer, 1926.

\bibitem{pauli-b}
W.~Pauli, ``Die allgemeinen prinzipien der wellenmechanik,'' in {\em
  Quantentheorie, Handbuch der Physik} (H.~Geiger and K.~Scheel, eds.),
  pp.~83--272, Berlin, Heidelberg: Springer, 1933.

\bibitem{galapon}
E.~Galapon, ``Self-adjoint time operator is the rule for discrete semi-bounded
  hamiltonians,'' {\em Proc. R. Soc. Lond. A}, vol.~458, pp.~2671--2689, 2002.

\bibitem{th01}
L.~Loveridge and T.~Miyadera, ``Relative quantum time,'' {\em Found. Phys.},
  vol.~49, pp.~549--560, 2019.

\bibitem{th02}
M.~Vogl, P.~Laurell, A.~Barr, and G.~Fiete, ``Analogue of hamilton-jacobi
  theory for the time-evolution operator,'' {\em Phys. Rev. A}, vol.~100,
  p.~012132, 2019.

\bibitem{th03}
J.~Ashmead, ``Time dispersion and quantum mechanics,'' {\em Phys. Conf. Ser.},
  vol.~1239, p.~012015, 2019.

\bibitem{th04}
A.~Schild, ``Time in quantum mechanics: A fresh look at the continuity
  equation,'' {\em Phys. Rev. A}, vol.~98, p.~052113, 2018.

\bibitem{th05}
N.~Argaman, ``A lenient causal arrow of time?,'' {\em Entropy}, vol.~20,
  p.~294, 2018.

\bibitem{th06}
A.~Smith and M.~Ahmadi, ``Quantizing time: Interacting clocks and systems,''
  {\em Quantum}, vol.~3, p.~160, 2019.

\bibitem{th07}
M.~Lienert, S.~Petrat, and R.~Tumulka, ``Multi-time wave functions versus
  multiple timelike dimensions,'' {\em Found. Phys.}, vol.~47, pp.~1582--1590,
  2017.

\bibitem{th08}
D.~Bruschi, ``Work drives time evolution,'' {\em Ann. Phys.}, vol.~394,
  pp.~155--161, 2018.

\bibitem{th09}
J.~Dressel, A.~Chantasri, A.~Jordan, and A.~Korotkov, ``Arrow of time for
  continuous quantum measurement,'' {\em Phys. Rev. Lett.}, vol.~119,
  p.~220507, 2017.

\bibitem{th10}
S.~Khorasani, ``Time operator in relativistic quantum mechanics,'' {\em Commun.
  Theor. Phys.}, vol.~68, pp.~35--38, 2017.

\bibitem{th11}
H.~Kitada, J.~Jeknic-Dugic, M.~Arsenijevic, and M.~Dugic, ``A minimalist
  approach to conceptualization of time in quantum theory,'' {\em Phys. Lett.
  A}, vol.~380, p.~3970, 2016.

\bibitem{th12}
P.~Aniello, F.~Ciaglia, F.~Di~Cosmo, G.~Marmo, and J.~P\'erez-Pardo, ``Time,
  classical and quantum,'' {\em Ann. Phys.}, vol.~373, pp.~532--543, 2016.

\bibitem{th13}
E.~Dias and F.~Parisio, ``Space-time symmetric extension of non-relativistic
  quantum mechanics,'' {\em Phys. Rev. A}, vol.~95, p.~032133, 2017.

\bibitem{th14}
A.~Sudbery, ``Time, chance and quantum theory,'' in {\em Probing the Meaning
  and Structure of Quantum Mechanics: Superpositions, Semantics, Dynamics and
  Identity} (D.~Aerts, C.~de~Ronde, H.~Freytes, and R.~Giuntini, eds.),
  pp.~324--339, Singapore: World Scientific, 2017.

\bibitem{th15}
V.~Overbeck and H.~Weimer, ``Time evolution of open quantum many-body
  systems,'' {\em Phys. Rev. A}, vol.~93, p.~012106, 2016.

\bibitem{th16}
S.~Banerjee, S.~Bera, and T.~Singh, ``Cosmological constant, quantum
  measurement, and the problem of time,'' {\em Int. J. Mod. Phys.}, vol.~24,
  p.~1544011, 2015.

\bibitem{th17}
T.~Miyadera, ``Energy-time uncertainty relations in quantum measurements,''
  {\em Found. Phys.}, vol.~46, pp.~1522--1550, 2016.

\bibitem{th18}
V.~Giovannetti, S.~Lloyd, and L.~Maccone, ``Quantum time,'' {\em Phys. Rev. D},
  vol.~92, p.~045033, 2015.

\bibitem{th19}
J.~Briggs, ``The equivalent emergence of time dependence in classical and
  quantum mechanics,'' {\em Phys. Rev. A}, vol.~91, p.~052119, 2015.

\bibitem{th20}
V.~Olkhovsky and E.~Recami, ``Time as a quantum observable,'' {\em Int. J. Mod.
  Phys. A}, vol.~22, pp.~5063--5087, 2007.

\bibitem{th21}
J.~Jing and H.~Ma, ``Polynomial scheme for time evolution of open and closed
  quantum systems,'' {\em Phys. Rev. E}, vol.~75, p.~016701, 2007.

\bibitem{th22}
R.~de~la Madrid and J.~Isidro, ``A selfadjoint variant of the time operator,''
  {\em Adv. Studies Theor. Phys.}, vol.~2, pp.~281--289, 2008.

\bibitem{th-pre01}
D.~Geiger and Z.~Kedem, ``A theory for time arrow.'' arXiv:1906.11712v2
  [quant-ph], 2019.

\bibitem{th-pre02}
D.~Buchholz and K.~Fredenhagen, ``Classical dynamics, arrow of time, and
  genesis of the heisenberg. commutation relations.'' arXiv:1905.02711
  [quant-ph], 2019.

\bibitem{th-pre03}
K.~Bryan and A.~Medved, ``The problem with `the problem of time'.''
  arXiv:1811.09660 [quant-ph], 2018.

\bibitem{th-pre04}
M.~Bauer, ``The problem of time in quantum mechanics.'' arXiv:1606.02618v2
  [quant-ph], 2016.

\bibitem{th-pre05}
F.~Dias, E.O. and~Parisio, ``Elements of a new approach to time in quantum
  mechanics.'' arXiv:1507.02899 [quant-ph], 2015.

\bibitem{th-pre06}
G.~Bacciagaluppi, ``Probability, arrow of time and decoherence.''
  arXiv:quant-ph/0701225, 2007.

\bibitem{th-rel-mech-a}
P.~H\"ohn, A.~Smith, and M.~Lock, ``Equivalence of approaches to relational
  quantum dynamics in relativistic settings.'' arXiv:2007.00580v1 [gr-qc],
  2020.

\bibitem{th-rel-mech-b}
P.~H\"ohn, A.~Smith, and M.~Lock, ``The trinity of relational quantum
  dynamics.'' arXiv:1912.00033v2 [quant-ph], 2020.

\bibitem{th-montevideo}
R.~Gambini and J.~Pullin, ``The montevideo interpretation: How the inclusion of
  a quantum gravitational notion of time solves the measurement problem,'' {\em
  Universe}, vol.~6, p.~236, 2020.

\bibitem{toa-th}
E.~Galapon and J.~Magadan, ``Quantizations of the classical time of arrival and
  their dynamics,'' {\em Ann. Phys.}, vol.~397, pp.~278--302, 2018.

\bibitem{toa-exp}
R.~Ximenes, F.~Parisio, and E.~Dias, ``Comparing experiments on quantum
  traversal time with the predictions of a space-time-symmetric formalism,''
  {\em Phys. Rev. A}, vol.~98, p.~032105, 2018.

\bibitem{termo-th01}
A.~Klimenko, ``The direction of time and boltzmann's time hypothesis,'' {\em
  Phys. Scr.}, vol.~94, p.~034002, 2019.

\bibitem{termo-th02}
W.~Wreszinski, ``Irreversibility, the time arrow and a dynamical proof of the
  second law of thermodynamics.'' arXiv:1902.07591v6 [math-ph], 2019.

\bibitem{termo-exp}
K.~Micadei, J.~Peterson, A.~Souza, R.~Sarthour, I.~Oliveira, G.~Landi,
  T.~Batalhao, R.~Serra, and E.~Lutz, ``Reversing the direction of heat flow
  using quantum correlations,'' {\em Nature Comm.}, vol.~10, p.~2456, 2019.

\bibitem{ent-th01}
M.~Nowakowski, ``Quantum entanglement in time,'' {\em AIP Conf. Proc.},
  vol.~1841, p.~020007, 2017.

\bibitem{ent-th02}
M.~Nowakowski, ``Monogamy of quantum entanglement in time.'' arXiv:1604.03976
  [quant-ph], 2016.

\bibitem{ent-exp}
E.~Moreva, G.~Brida, M.~Gramegna, V.~Giovannetti, L.~Maccone, and M.~Genovese,
  ``Time from quantum entanglement: an experimental illustration,'' {\em Phys.
  Rev. A}, vol.~89, p.~052122, 2014.

\bibitem{qcomp01}
A.~Grimsmo, ``Time-delayed quantum feedback control,'' {\em Phys. Rev. Lett.},
  vol.~115, p.~060402, 2015.

\bibitem{qcomp02}
S.~Mamataj, D.~Saha, and N.~Banu, ``A review of reversible gates and its
  application in logic design,'' {\em AJER}, vol.~3, pp.~151--161, 2014.

\bibitem{qcomp03}
T.~Koike and Y.~Okudaira, ``Time complexity and gate complexity,'' {\em Phys.
  Rev. A}, vol.~82, p.~042305, 2010.

\bibitem{qcomp04}
J.~Rice, ``An introduction to reversible latches,'' {\em Comput. J.}, vol.~51,
  pp.~700--709, 2008.

\bibitem{qneural}
A.~Dendukuri, B.~Keeling, A.~Fereidouni, J.~Burbridge, K.~Luu, and
  H.~Churchill, ``Defining quantum neural networks via quantum time
  evolution.'' arXiv:1905.10912v2 [cs.LG], 2020.

\bibitem{tmeas}
V.~Belavkin and M.~Perkins, ``The nondemolition measurement of quantum time,''
  {\em Int. J. Theor. Phys.}, vol.~37, pp.~219--226, 1998.

\bibitem{tmeas-cosmo}
N.~Kajuri, ``The time measurement problem in quantum cosmology,'' {\em Int. J.
  Mod. Phys. D}, vol.~26, p.~1743011, 2017.

\bibitem{dmat}
E.~Alvarez, ``Exercise: Dark matter as fields that evolve backward in time.''
  arXiv:1803.08531 [gr-qc], 2018.

\bibitem{wheeler-a}
J.~Wheeler, ``The ``past'' and the ``delayed-choice'' double-slit experiment,''
  in {\em Mathematical Foundations of Quantum Theory} (A.~Marlow, ed.),
  pp.~9--48, New York, USA: Academic Press, 1978.

\bibitem{wheeler-b}
J.~Wheeler, ``Law without law,'' in {\em Quantum Theory and Measurement}
  (J.~Wheeler and Z.~W.H., eds.), pp.~182--213, Princeton, USA: Princeton
  University Press, 1984.

\bibitem{exp-aspect}
A.~Aspect, J.~Dalibard, and G.~Roger, ``Experimental test of bell's
  inequalities using time-varying analysers,'' {\em Phys. Rev. Lett.}, vol.~49,
  pp.~1804--1807, 1982.

\bibitem{exp-wh-a}
A.~Zajonc, L.~Wang, X.~Zou, and L.~Mandel, ``Quantum eraser,'' {\em Nature},
  vol.~353, pp.~507--508, 1991.

\bibitem{exp-wh-b}
P.~Kwiat, A.~Steinberg, and R.~Chiao, ``Three proposed ``quantum erasers'',''
  {\em Phys. Rev. A}, vol.~49, pp.~61--68, 1994.

\bibitem{exp-wh-c}
T.~Herzog, P.~Kwiat, H.~Weinfurter, and A.~Zeilinger, ``Complementarity and the
  quantum eraser,'' {\em Phys. Rev. Lett.}, vol.~75, pp.~3034--3037, 1995.

\bibitem{exp-wh-d}
T.~Pittman, D.~Strekalov, A.~Migdall, M.~Rubin, A.~Sergienko, and Y.~Shih,
  ``Can two-photon interference be considered the interference of two
  photons?,'' {\em Phys. Rev. Lett.}, vol.~77, pp.~1917--1920, 1996.

\bibitem{exp-wh-e}
V.~Jacques, E.~Wu, F.~Grosshans, T.~F., P.~Grangier, A.~Aspect, and J.~Roch,
  ``Experimental realization of wheeler's delayed-choice gedanken experiment,''
  {\em Science}, vol.~315, pp.~966--968, 2007.

\bibitem{manning}
A.~Manning, R.~Khakimov, R.~Dall, and A.~Truscott, ``Wheeler's delayed-choice
  gedanken experiment with a single atom,'' {\em Nature Physics}, vol.~11,
  pp.~539--542, 2015.

\bibitem{exp-eraser}
Y.~Ho~Kim, R.~Yu, S.~Kulik, Y.~Shih, and M.~Scully, ``Delayed ``choice''
  quantum eraser,'' {\em Phys. Rev. Lett.}, vol.~84, pp.~1--5, 2000.

\bibitem{exp-space}
F.~Vedovato, C.~Agnesi, M.~Schiavon, D.~Dequal, L.~Calderaro, M.~Tomasin,
  D.~Marangon, A.~Stanco, V.~Luceri, G.~Bianco, G.~Vallone, and P.~Villoresi,
  ``Extending wheeler's delayed-choice experiment to space,'' {\em Sci. Adv.},
  vol.~3, p.~e1701180, 2017.

\bibitem{exp-lapalma-a}
R.~Ursin, F.~Tiefenbacher, T.~Schmitt-Manderbach, H.~Weier, T.~Scheidl,
  M.~Lindenthal, B.~Blauensteiner, T.~Jennewein, J.~Perdigues, P.~Trojek,
  B.~\"Omer, M.~F\"urst, M.~Meyenburg, J.~Rarity, Z.~Sodnik, C.~Barbieri,
  H.~Weinfurter, and A.~Zeilinger, ``Entanglement-based quantum communication
  over 144 km,'' {\em Nature Physics}, vol.~3, pp.~481--486, 2007.

\bibitem{exp-lapalma-b}
X.-S. Ma, J.~Kofler, A.~Qarry, N.~Tetik, T.~Scheidl, R.~Ursin, S.~Ramelow,
  T.~Herbst, L.~Ratschbacher, A.~Fedrizzi, T.~Jennewein, and A.~Zeilinger,
  ``Quantum erasure with causally disconnected choice,'' {\em PNAS}, vol.~110,
  pp.~1221--1226, 2013.

\bibitem{exp-2slits-a}
U.~Houser, W.~Neuwirth, and N.~Thesen, ``Time-dependent modulation of the
  probability amplitude of single photons,'' {\em Phys. Lett. A}, vol.~49,
  pp.~57--58, 1974.

\bibitem{th-2slits-a}
P.~Busch, ``On the energy-time uncertainty relation. part i: Dynamical time and
  time indeterminacy,'' {\em Found. Phys.}, vol.~20, pp.~1--32, 1990.

\bibitem{th-2slits-b}
P.~Busch, ``The time-energy uncertainty relation,'' in {\em Time in Quantum
  Mechanics} (J.~Muga, R.~Sala~Mayato, and I.~Egusqiza, eds.), pp.~73--105,
  Berlin Heidelberg: Springer, 2002.

\bibitem{exp-2slits-b}
F.~Lindner, M.~Sch\"atzel, H.~Walther, A.~Baltu{\v s}ka, E.~Goulielmakis,
  F.~Krausz, D.~Milo{\v s}evi\'c, D.~Bauer, W.~Becker, and G.~Paulus,
  ``Attosecond double-slit experiment,'' {\em Phys. Rev. Lett.}, vol.~95,
  p.~040401, 2005.

\bibitem{time-ent}
E.~Moreva, G.~Brida, M.~Gramegna, V.~Giovannetti, L.~Maccone, and M.~Genovese,
  ``Time from quantum entanglement: An experimental illustration,'' {\em Phys.
  Rev. A}, vol.~89, p.~052122, 2014.

\bibitem{emch1972}
G.~Emch, {\em Algebraic Methods in Statistical Mechanics and Quantum Field
  Theory}.
\newblock New York, USA: Wiley--Interscience, A Division of John Wiley \& Sons,
  Inc., 1972.

\bibitem{Landsman}
N.~Landsman, ``Algebraic quantum mechanics.''
  https://www.math.ru.nl/$\sim$landsman/algebraicQM.pdf.

\bibitem{Beny2020}
C.~B\'eny and F.~Richter, ``Algebraic approach to quantum theory: a
  finite-dimensional guide.'' arXiv:1505.03106v8 [quant-ph], 2020.

\bibitem{Busch1996}
P.~Busch, P.~Lahti, and P.~Mittelstaedt, {\em The Quantum Theory of Measurement
  (2 ed.)}.
\newblock Springer, 1996.

\bibitem{Lueders}
G.~L\"uders, ``Concerning the state-change due to the measurement process,''
  {\em Ann.~Phys. (Leipzig)}, vol.~8, p.~322, 1951.
\newblock reprinted in: Ann. Phys. (Leipzig) 15, 663 (2006).

\bibitem{Krauss}
K.~Krauss, {\em States, Effects and Operations: Fundamental Notions of Quantum
  Theory}.
\newblock Springer, 1983.

\bibitem{Choi}
M.~Choi, ``Completely positive linear maps on complex matrices,'' {\em Lin.
  Alg. App.}, vol.~10, pp.~285--290, 1975.

\bibitem{Ulmann1976}
A.~Uhlmann, ``The ``transition probability'' in the state space of a
  *-algebra,'' {\em Rep. Math. Phys.}, vol.~9, pp.~273--279, 1976.

\bibitem{Ulmann2016}
A.~Uhlmann, ``Transition probability (fidelity) and its relatives.''
  arXiv:1106.0979v2 [quant-ph], 2016.

\bibitem{GPPSchwarz2022}
A.~G\'o\'zd\'z, A.~P\c{e}drak, and W.~Piechocki, ``Ascribing quantum system to
  schwarzschild spacetime with naked singularity,'' {\em Class. Quantum Grav.},
  vol.~39, p.~145005, 2022.

\bibitem{GPPBKL2022-a}
A.~G\'o\'zd\'z, A.~P\c{e}drak, and W.~Piechocki, ``Quantum dynamics
  corresponding to the chaotic bkl scenario,'' {\em Eur. Phys. J. C}, vol.~83,
  p.~150, 2023.

\bibitem{GPPBKL2022-b}
A.~G\'o\'zd\'z, A.~P\c{e}drak, and W.~Piechocki, ``Quantum dynamics
  corresponding to the classical bkl scenario.'' arXiv:2204.11274v1 [gr-qc],
  2022.

\bibitem{Gazeau2015}
J.~Gazeau and B.~Heller, ``Positive--operator valued measure (povm)
  quantization,'' {\em Axioms}, vol.~4, pp.~1--29, 2015.

\bibitem{epr-speed}
J.~Yin, Y.~Cao, H.-L. Yong, J.-G. Ren, H.~Liang, S.-K. Liao, F.~Zhou, C.~Liu,
  Y.-P. Wu, G.-S. Pan, L.~Li, N.-L. Liu, Q.~Zhang, C.-Z. Peng, and J.-W. Pan,
  ``Lower bound on the speed of nonlocal correlations without locality and
  measurement choice loopholes,'' {\em Phys. Rev. Lett.}, vol.~110, p.~260407,
  2013.

\bibitem{Debicki2007}
A.~G\'{o}\'{z}d\'{z} and M.~D\c{e}bicki, ``Time operator and quantum projection
  evolution,'' {\em Phys. Atom. Nucl.}, vol.~70, pp.~529--536, 2007.

\bibitem{Robertson1929}
H.~Robertson, ``The uncertainty principle,'' {\em Phys. Rev.}, vol.~34, p.~163,
  1929.

\bibitem{muga2}
J.~Muga, A.~Ruschhaupt, and A.~Del~Campo, eds., {\em Time in Quantum Mechanics
  -- vol.2}.
\newblock Berlin, Heidelberg: Springer, 2009.

\bibitem{Cov-Hohn2018}
P.~H\"ohn and A.~Vanrietvelde, ``How to switch between relational quantum
  clocks.'' arXiv:1810.04153 [gr-qc], 2018.

\bibitem{Cov-Hohn2019}
P.~H\"ohn, ``Switching internal times and a new perspective on the `wave
  function of the universe','' {\em Universe}, vol.~5, p.~116, 2019.

\bibitem{Cov-Giacomini2019}
F.~Giacomini, E.~Castro-Ruiz, and {\v C}.~Brukner, ``Quantum mechanics and the
  covariance of physical laws in quantum reference frames,'' {\em Nat.
  Commun.}, vol.~10, p.~494, 2019.

\bibitem{Cov-Vanrietvelde2018}
A.~Vanrietvelde, P.~H\"ohn, and F.~Giacomini, ``Switching quantum reference
  frames in the n-body problem and the absence of global relational
  perspectives.'' arXiv:1809.05093 [quant-ph], 2018.

\bibitem{Cov-Vanrietvelde2020}
A.~Vanrietvelde, P.~H\"ohn, F.~Giacomini, and E.~Castro-Ruiz, ``A change of
  perspective: switching quantum reference frames via a perspective-neutral
  framework,'' {\em Quantum}, vol.~4, p.~225, 2020.

\bibitem{Nonloc-CastroRuiz2020}
E.~Castro-Ruiz, F.~Giacomini, A.~Belenchia, and {\v C}.~Brukner, ``Quantum
  clocks and the temporal localisability of events in the presence of
  gravitating quantum systems,'' {\em Nat. Commun.}, vol.~11, p.~2672, 2020.

\bibitem{Gozdz2015}
A.~G\'{o}\'{z}d\'{z}, K.~Rybak, A.~P\c{e}drak, and M.~G\'{o}\'{z}d\'{z},
  ``Quantum time in nuclear physics,'' {\em Acta Phys. Pol. B Proc. Supp.},
  vol.~8, pp.~591--596, 2015.

\bibitem{DealyedChoice2008}
A.~G\'o\'zd\'z and K.~Stefa\'nska, ``Projection evolution and delayed--choice
  experiments,'' {\em J. Phys.: Conf. Ser.}, vol.~104, p.~012007, 2008.

\bibitem{DelayedChoice2018}
M.~G\'o\'zd\'z, A.~G\'o\'zd\'z, A.~Gusev, and S.~Vinitsky, ``Projection
  evolution of quantum states-the delayed choice puzzle,'' {\em Phys. Atom.
  Nucl.}, vol.~81, pp.~853--857, 2018.

\bibitem{Gozdz2014}
M.~G\'o\'zd\'z and A.~G\'o\'zd\'z, ``On particle oscillations,'' {\em Phys.
  Scr.}, vol.~89, p.~054010, 2014.

\end{thebibliography}

%%%%%%%%%%%%%%%%%%%%%%%%%%%%%%%%%%%%%%%%%%%%%%%%%%%%%%%%%%%%%%%%%%%%%%% 

%%% BibTeX
%%%%%%%%%%%%%%%%%%%%%%%%%%%%%%%%%%%%%%%%%%%%%%%%%%%%%%%%%%%%%%%%%%%%%%%
%\bibliographystyle{plain}
%\bibliography{PEv}
%%%%%%%%%%%%%%%%%%%%%%%%%%%%%%%%%%%%%%%%%%%%%%%%%%%%%%%%%%%%%%

\end{document}